\documentclass[aps,prb,twocolumn,showpacs,superscriptaddress,floatfix]{revtex4}
\usepackage{bm}
\usepackage{color}
\usepackage{amssymb}
\usepackage{graphicx}
\usepackage{amsmath}
\usepackage{dcolumn}
\usepackage{bm}
\usepackage{soul}
\usepackage[german,english]{babel}

\begin{document}

\title{Screening and plasmons in pure and disordered single- and bilayer black phosphorus}
\date{\today }
\author{Fengping Jin}
\affiliation{Institute for Advanced Simulation, J\"ulich Supercomputing Centre,\\
Forschungszentrum J\"ulich, D-52425 J\"ulich, Germany}
\author{Rafael Rold\'an}
\email{rafael.roldan.toro@gmail.com}
\affiliation{Instituto Madrile\~no de Estudios Avanzados en Nanociencia (IMDEA-Nanociencia), 28049 Madrid, Spain}
\author{Mikhail I. Katsnelson}
\affiliation{Radboud University, Institute for Molecules and Materials,
Heijendaalseweg 135, 6525 AJ Nijmegen, The Netherlands}
\author{Shengjun Yuan}
\email{s.yuan@science.ru.nl}
\affiliation{Radboud University, Institute for Molecules and Materials,
Heijendaalseweg 135, 6525 AJ Nijmegen, The Netherlands}

\begin{abstract}
We study collective plasmon excitations and screening of pure and disordered single-
and bilayer black phosphorus beyond the low energy continuum approximation.
The dynamical polarizability of phosphorene is computed using a
tight-binding model that properly accounts for the band structure in a wide
energy range. Electron-electron interaction is considered within the Random
Phase Approximation. Damping of the plasmon modes due to different kinds of
disorder, such as resonant scatterers and long-range disorder potentials, is
analyzed. We further show that an electric field applied perpendicular to
bilayer phosphorene can be used to tune the dispersion of the plasmon modes.
For sufficiently large electric field, the bilayer BP enters in a
topological phase with a characteristic plasmon spectrum, which is gaped in the
armchair direction.
\end{abstract}

\maketitle

\section{Introduction}

Phosphorene is a new kind of two-dimensional (2D) material that can be
obtained by mechanical exfoliation method from black phosphorus (BP) films%
\cite{LLi2014,HLiu2014,XiaF2014,Koenig2014,Castellanos2014,LiL2014}. It is a
semiconductor with direct band gap and highly anisotropic electronic and
optical properties\cite%
{Morita1986,Ling2015,Qiao2014,Rudenko2014,GuanJ2014,PengXH2014,XiaF2014,Tran2014,Cakir2014,Low2014,Low2014b,LiPK2014,Yuan2015BP}%
. The band gap of thin BP films varies, depending on the thickness, from 0.3
eV to around 2 eV. Furthermore, the electronic band structure of this
material is very sensitive to strain,\cite%
{Rodin2014,HLiu2014,PengXH2014,Elahi2015,RG15} making it a promising
candidate for electromechanical applications. The plasmons and screening in
pristine single-layer and multilayer phosphorene were studied theoretically
in Ref. \onlinecite{Low2014p} within a low energy $\mathbf{k\cdot p}$ model,
where it was found that the dispersion of the collective modes in this
material strongly depends on the direction of propagation. A similar
approximation was later used in Ref. \onlinecite{Rodin2015} to study a
generic double layer anisotropic system with large interlayer distance.
Collective excitations of BP in the presence of a quantizying magnetic field
have been studied recently, finding that the excitation spectrum is
discretized into a series of anisotropic magneto-excitons.\cite{Jiang2015}

A fundamental issue that needs to be addressed is the effect of disorder on
the electronic and optical properties of this material. Phosphorene and its
layered structures are highly sensitive to the environment\cite%
{Castellanos2014,Favron2014,Wood2014,Island2015}, due to its high reactivity
when the samples are exposed to air. In this paper, we study the plasmon
modes of disordered single- and bilayer BP by using a tight-binding (TB)
model which properly accounts for the electronic band dispersion in a wide
energy window.\cite{Rudenko2014} Damping of plasmons due to different kinds
of disorder, such as point defects (resonant scatterers) and long-range
disorder potential, is analyzed. For this aim, the polarization function is
calculated with the tight-binding propagation method (TBPM) \cite%
{YRK10,Yuan2011,Yuan2012}, which is extremely efficient in large-scale
calculations of systems with more than millions of atoms. The dielectric
function is obtained within the random phase approximation (RPA). For the
case of bilayer BP, we study the effect of an electric field applied
perpendicular to the layers, which can be used to tune the dispersion of the
plasmon modes. Finally we discuss the excitation spectrum of bilayer BP in the topological phase driven by the application of bias, in which the band structure is gapped in the armchair direction and dispersive in the zigzag direction. 

The paper is organized as follows. In Sec. \ref{Sec:Method} we describe the method used in our calculations. The dielectric screening properties are discussed in Sec. \ref{Sec:Screening} and the plasmon excitation spectrum of disordered BP is analyzed in Sec. \ref{Sec:Plasmons}. Our main conclusions are summarized in Sec. \ref{Sec:Conclusions}.

\begin{figure}[tb]
\begin{center}
\includegraphics[width=8.25cm]{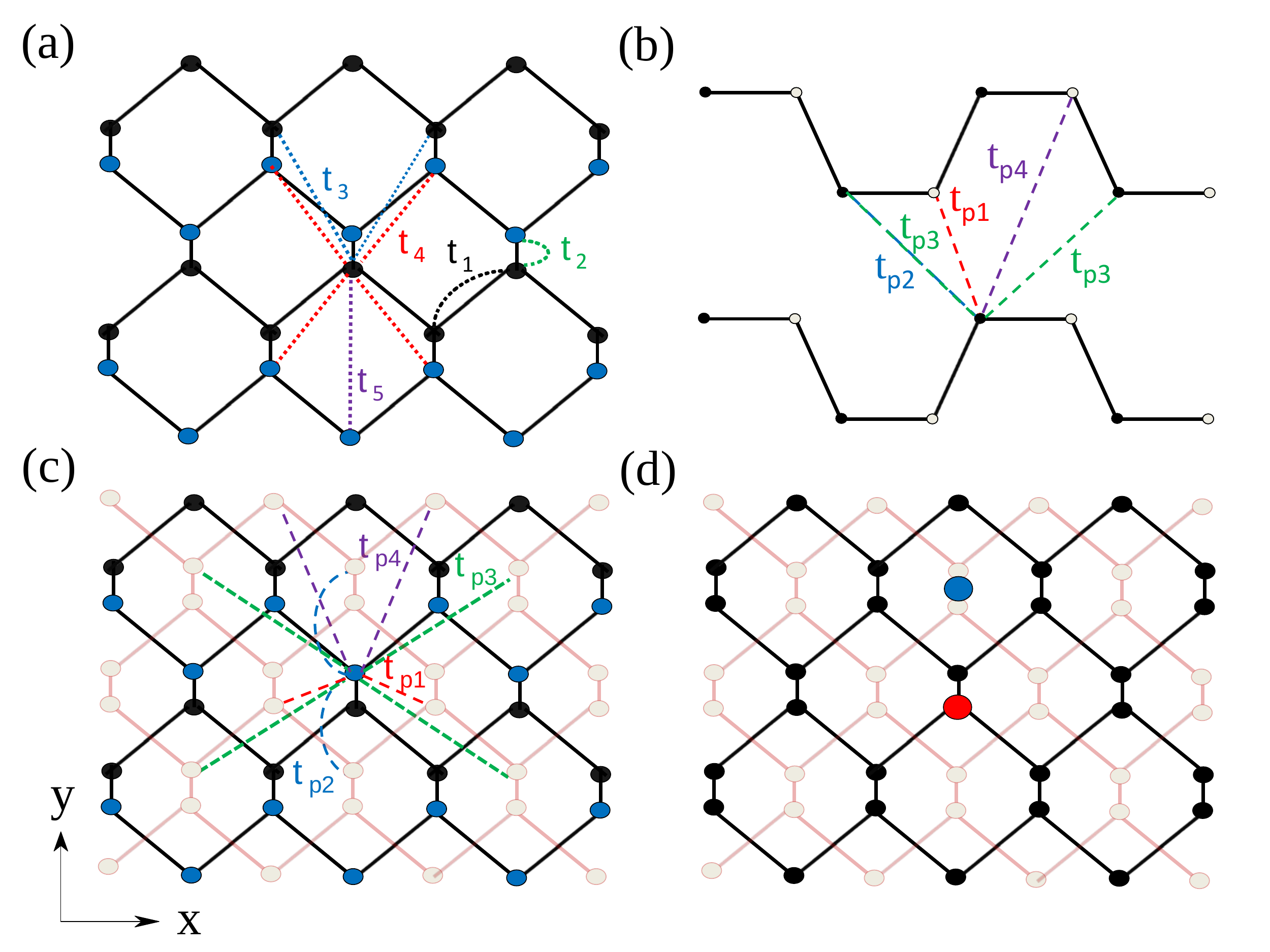}
\end{center}
\caption{Lattice structure of single-layer and bilayer BP. The relevant
hopping terms considered in the Hamiltonian (\protect\ref{Eq:Hamiltonian})
are shown. The red and blue dots in (d) represent a resonant scatterer and a
Gaussian center, respectively.}
\label{Fig:Lattice}
\end{figure}

\section{Model and Method}\label{Sec:Method}

BP is a single-elemental layered crystalline material consisting of only
phosphorus atoms arranged in a puckered orthorhombic lattice. As sketched in Fig. \ref{Fig:Lattice}, single layer
phosphorene contains two atomic layers and two kinds of bonds with 2.22 \AA %
~ and 2.24 \AA ~ bond lengths for in-plane and inter-plane P-P connections,
respectively.\cite{Morita1986} The electronic band structure around the gap
can be described with a TB Hamiltonian for pristine BP with the form\cite%
{Rudenko2014} 
\begin{equation}
\mathcal{H}=\sum_{i}\epsilon _{i}n_{i}+\sum_{i\neq j}t_{ij}c_{i}^{\dagger
}c_{j}+\sum_{i\neq j}t_{p,ij}c_{i}^{\dagger }c_{j},  \label{Eq:Hamiltonian}
\end{equation}%
where we consider five intralayer ($t_{1}=-1.220$ eV, $t_{2}=3.665$ eV, $%
t_{3}=-0.205$ eV, $t_{4}=-0.105$ eV, $t_{5}=-0.055$ eV) and four interlayer
hopping terms ($t_{p1}=0.295$ eV, $t_{p2}=0.273$ eV, $t_{p3}=-0.151$ eV, $%
t_{p4}=-0.091$ eV). For unbiased bilayer BP, there is an energy splitting of 
$\Delta \epsilon =1.0$ eV between the nonequivalent electrons in the
sublayers. This model, which is based on first principle $GW$ calculations,
accurately reproduces the conduction and valence bands for energies $\sim $%
0.3 eV beyond the gap. For a detailed description of the above TB
Hamiltonian we refer the reader to Ref. \onlinecite{Rudenko2014}.

We consider two main different sources of disorder in BP: local point
defects and long range disorder potential. Point defects are modeled by
elimination of atoms randomly over the whole sample, which can be viewed as
vacancies, chemical adsorbates or substitution of other types of atoms which
prevent the hopping of electrons to their neighbor sites.\cite%
{Neto2009,Sarma2011,Katsnelsonbook,YRK10,YRK10b,QiuH2013,Yuan2014,Yuan2015BP}
This type of point defects are the so called resonant scatterers which
provide resonances within the band gap \cite{Yuan2014,QiuH2013,Yuan2015BP},
as confirmed by first-principles calculations which include single vacancies%
\cite{Liu2014,HuW2014}, adatoms (Si, Ge, Au, Ti, V)\cite{Kulish2015},
absorption of organic molecules\cite{Zhang2014}, substitutional p-dopants
(Te, C)\cite{Liu2014} or oxygen bridge-type defects\cite{Ziletti2015}.

The long-range disorder potential (LRDP), on the other hand, can account for
electron-hole puddles, which are random changes of the local potential due
to the inhomogeneous distribution of charge in the sample. Within the TB
model, it can be considered as a correlated Gaussian potential,\cite%
{LMC08,YRK10b,YRRK11} such that the potential at site $i$ follows 
\begin{equation}
v_{i}=\sum_{k}U_{k}\exp \left( -\frac{\left\vert \mathbf{r}_{i}-\mathbf{r}%
_{k}\right\vert ^{2}}{2d^{2}}\right) ,  \label{vgaussian}
\end{equation}%
where $\mathbf{r}_{k}$ is the position of the $k$-th Gaussian center, which
are chosen to be randomly distributed over the centers of the projected
lattice on the surface, $U_{k}$ is the amplitude of the potential at the $k$%
-th Gaussian center, which is uniformly random in the range $[-\Delta
,\Delta ]$, and $d$ is interpreted as the effective potential radius.
Typical values of $\Delta $ and $d$ used in our model are $\Delta =5$~eV and $%
d=5a$ for LRDP.\cite{YRK10b,YRRK11,Yuan2015BP} Here $a\approx 2.216$\AA\ is
the atomic distance of two nearest neighbors within the same plane. The
origin of LRDP could be screened charged impurities on the substrate \cite%
{Hwang2007,ZhangYB2009,Rudenko2011,Principi13} or surface corrugations such
as ripples and wrinkles\cite{Gibertini2010,Gibertini2012}. Although LRDP
does not introduce resonances in the spectrum, they lead to an uniform
increase of states in the gap.\cite{Yuan2015BP} The amount of resonant
scatterers and Gaussian centers are quantified by $n_{x}$ and $n_{c}$,
respectively, which are the probability for a scatterer or Gaussian center
to exist.

The excitation spectrum and screening of single- and bilayer BP are
calculated by using the TBPM.\cite{YRK10,Yuan2011,Yuan2012} The dynamic
polarization function is obtained from the Kubo formula \cite{K57} 
\begin{equation}
\Pi \left( \mathbf{q},\omega \right) =\frac{i}{V}\int_{0}^{\infty }d\tau
e^{i\omega \tau }\left\langle \left[ \rho \left( \mathbf{q},\tau \right)
,\rho \left( -\mathbf{q},0\right) \right] \right\rangle ,  \label{Eq:Kubo}
\end{equation}%
where $V$ denotes the volume (or area in 2D) of the unit cell, $\rho \left( 
\mathbf{q}\right) $ is the density operator given by%
\begin{equation}
\rho \left( \mathbf{q}\right)
=\sum_{l=1}^{N_{layer}}\sum_{i}c_{l,i}^{\dagger }c_{l,i}\exp \left( i\mathbf{%
q\cdot r}_{l,i}\right) ,
\end{equation}%
and the average in (\ref{Eq:Kubo}) is taken over the canonical ensemble. For
the case of the single-particle Hamiltonian, Eq.~(\ref{Eq:Kubo}) can be
written as\cite{Yuan2011}%
\begin{eqnarray}
&&\Pi \left( \mathbf{q},\omega \right) =-\frac{2}{V}\int_{0}^{\infty }d\tau
e^{i\omega \tau }  \notag  \label{Eq:Kubo2} \\
&&\times \mathrm{Im}\left\langle \varphi \right\vert n_{F}\left( H\right)
e^{iH\tau }\rho \left( \mathbf{q}\right) e^{-iH\tau }\left[ 1-n_{F}\left(
H\right) \right] \rho \left( -\mathbf{q}\right) \left\vert \varphi
\right\rangle ,  \notag \\
&&
\end{eqnarray}%
where $n_{F}\left( H\right) =\frac{1}{e^{\beta \left( H-\mu \right) }+1}$ is
the Fermi-Dirac distribution operator, $\beta =1/k_{B}T$ where $k_{B}$ is
the Boltzmann constant and $T$ is the temperature, and $\mu $ is the
chemical potential. We use units such that $\hbar =1$ and the average in
Eq.~(\ref{Eq:Kubo2}) is performed over a random phase superposition of all
the basis states in the real space, i.e.,\cite{HR00,YRK10} 
\begin{equation}
\left\vert \varphi \right\rangle =\sum_{l,i}a_{l,i}c_{l,i}^{\dagger
}\left\vert 0\right\rangle ,  \label{Eq:phi0}
\end{equation}%
where the coefficients $a_{l,i}$ are random complex numbers normalized as $%
\sum_{l,i}\left\vert a_{l,i}\right\vert ^{2}=1$. We next introduce the time
evolution of two wave functions 
\begin{eqnarray}
\left\vert \varphi _{1}\left( \mathbf{q,}\tau \right) \right\rangle
&=&e^{-iH\tau }\left[ 1-n_{F}\left( H\right) \right] \rho \left( -\mathbf{q}%
\right) \left\vert \varphi \right\rangle , \\
\left\vert \varphi _{2}\left( \tau \right) \right\rangle &=&e^{-iH\tau
}n_{F}\left( H\right) \left\vert \varphi \right\rangle ,
\end{eqnarray}%
which allows us to express the real and imaginary part of the dynamic
polarization as 
\begin{eqnarray}
\mathrm{Re}\Pi \left( \mathbf{q},\omega \right) &=&-\frac{2}{V}%
\int_{0}^{\infty }d\tau \cos (\omega \tau )~\text{Im}\left\langle \varphi
_{2}\left( \tau \right) \left\vert \rho \left( \mathbf{q}\right) \right\vert
\varphi _{1}\left( \tau \right) \right\rangle ,  \notag  \label{Eq:RePi-ImPi}
\\
\mathrm{Im}\Pi \left( \mathbf{q},\omega \right) &=&-\frac{2}{V}%
\int_{0}^{\infty }d\tau \sin (\omega \tau )~\text{Im}\left\langle \varphi
_{2}\left( \tau \right) \left\vert \rho \left( \mathbf{q}\right) \right\vert
\varphi _{1}\left( \tau \right) \right\rangle .  \notag \\
&&
\end{eqnarray}%
The time evolution operator $e^{-iH\tau }$ and Fermi-Dirac distribution
operator $n_{F}\left( H\right) $ can be obtained from the standard Chebyshev
polynomial decomposition.\cite{YRK10} In our calculations, the chemical
potential is set to be $\sim 0.1$~eV above the edge of conduction band, i.e., 
$\mu =0.4$~eV for single-layer and $0.73$~eV for bilayer BP. The temperature
is fixed at $T=300$~K. We use periodic boundary conditions, and the system
size is $8192\times 8192$ for single-layer BP, and $6000\times 6000$ for
bilayer.

\begin{figure}[tb]
\begin{center}
\mbox{
\includegraphics[width=4.25cm]{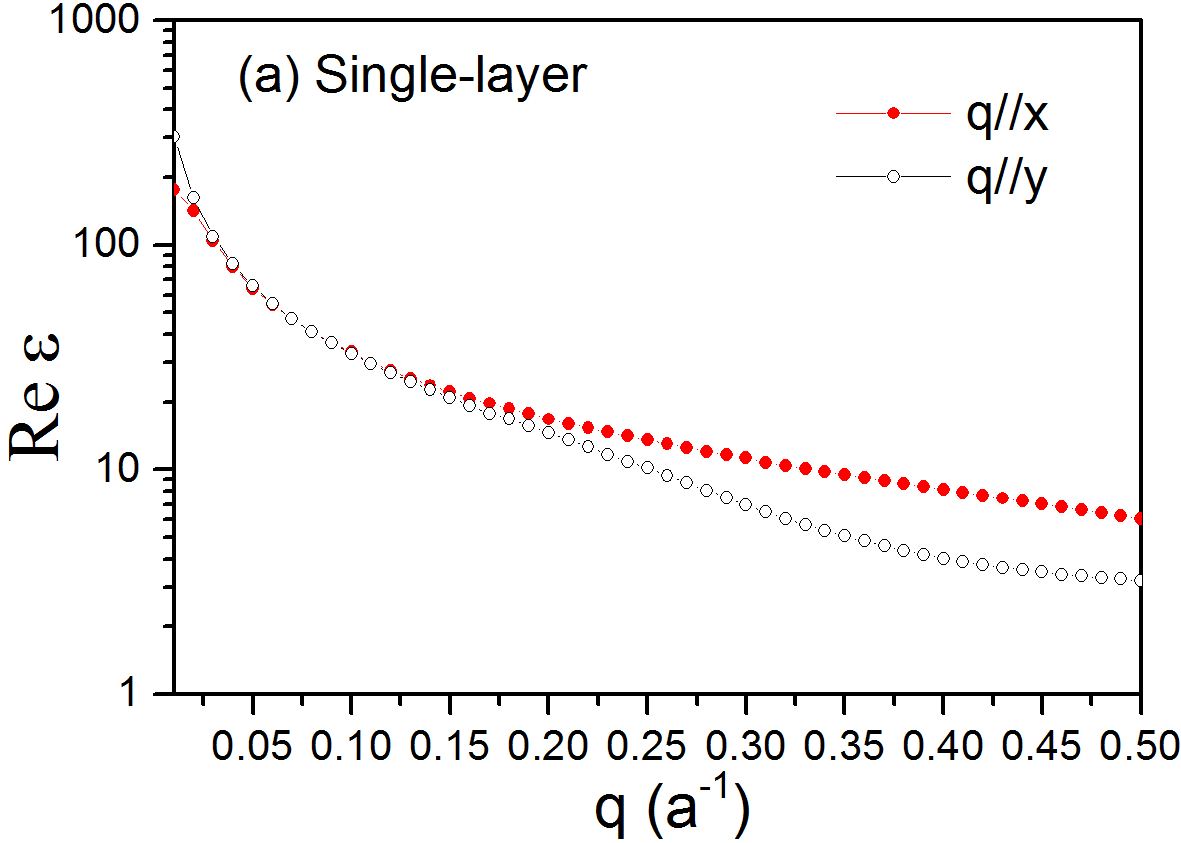}
\includegraphics[width=4.25cm]{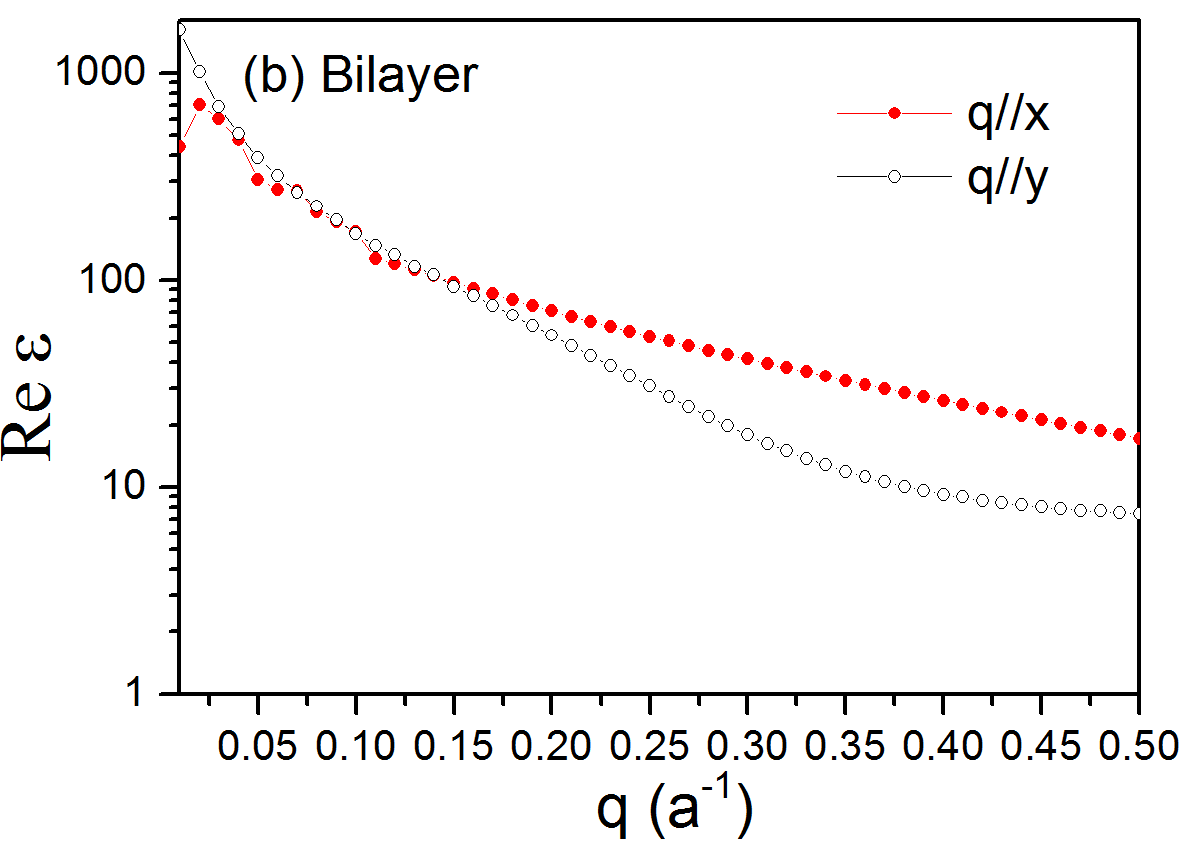}
} 
\mbox{
\includegraphics[width=4.25cm]{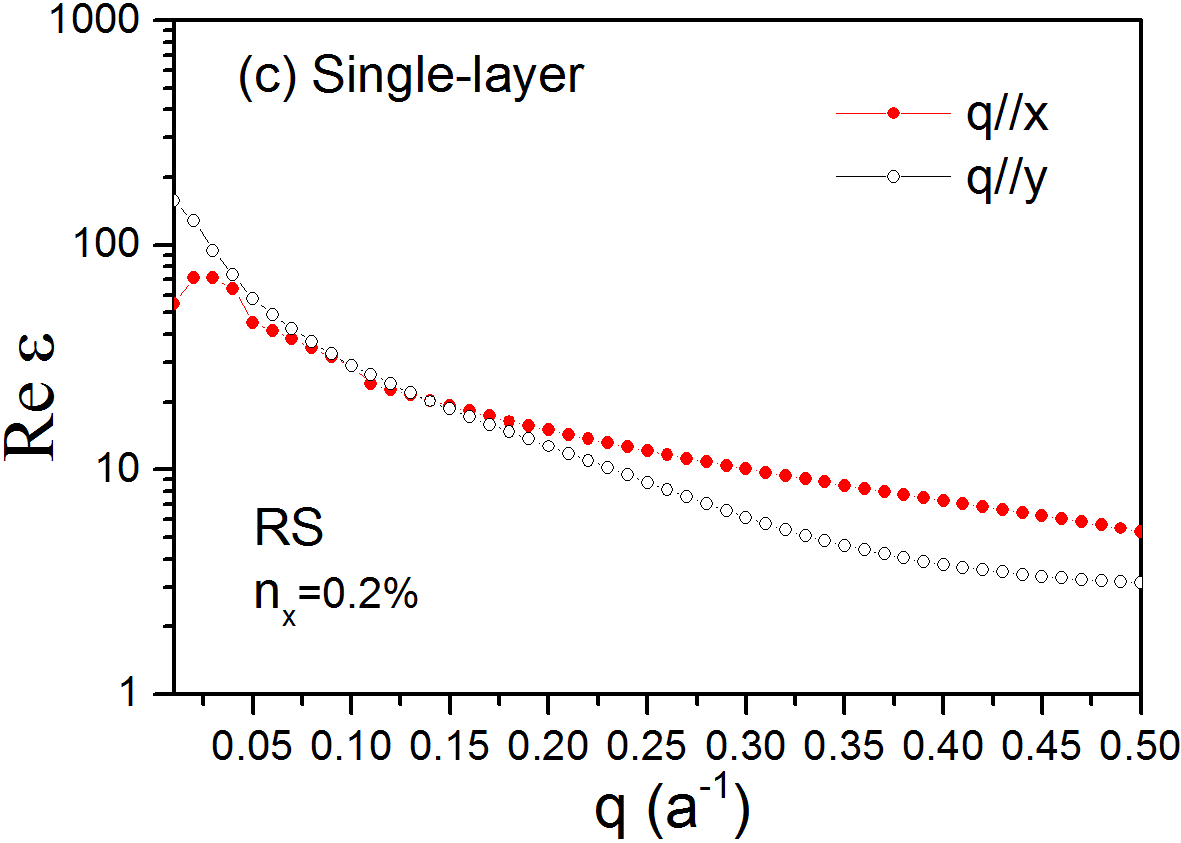}
\includegraphics[width=4.25cm]{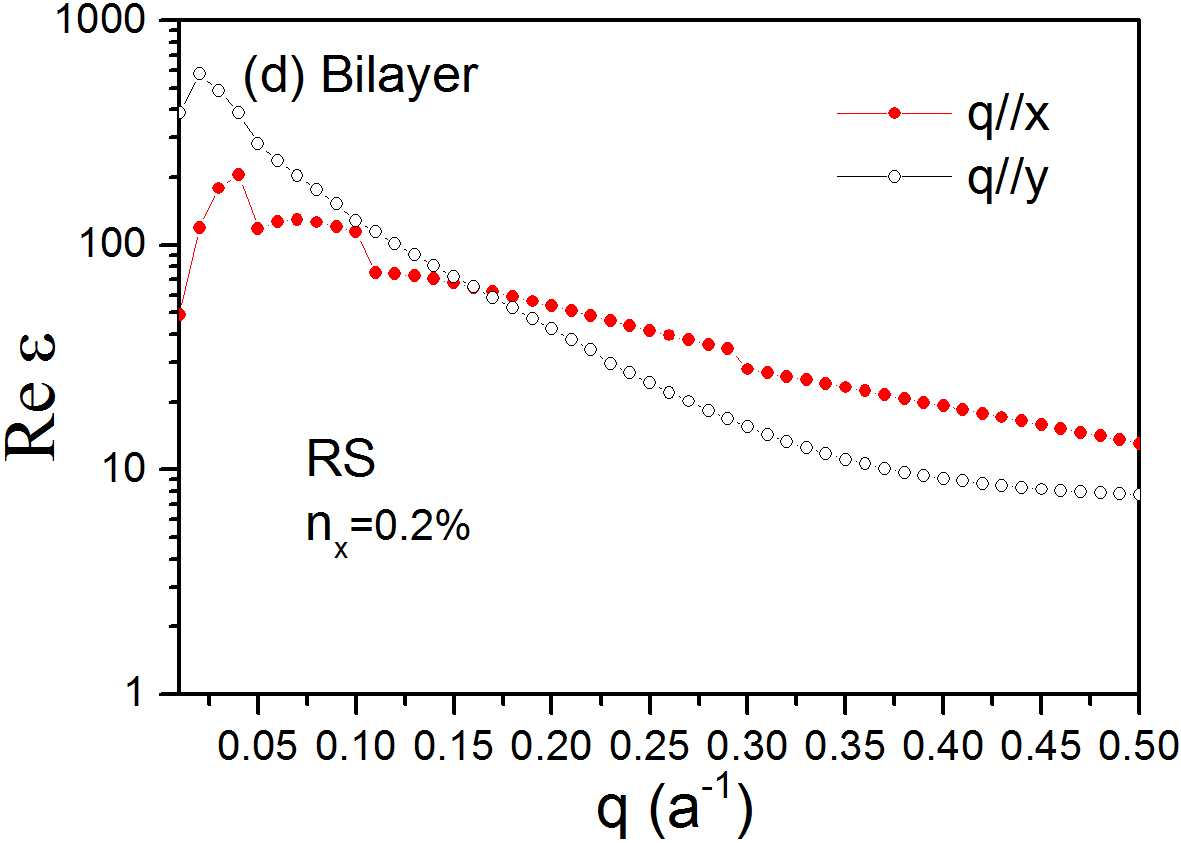}
} \includegraphics[width=4.25cm]{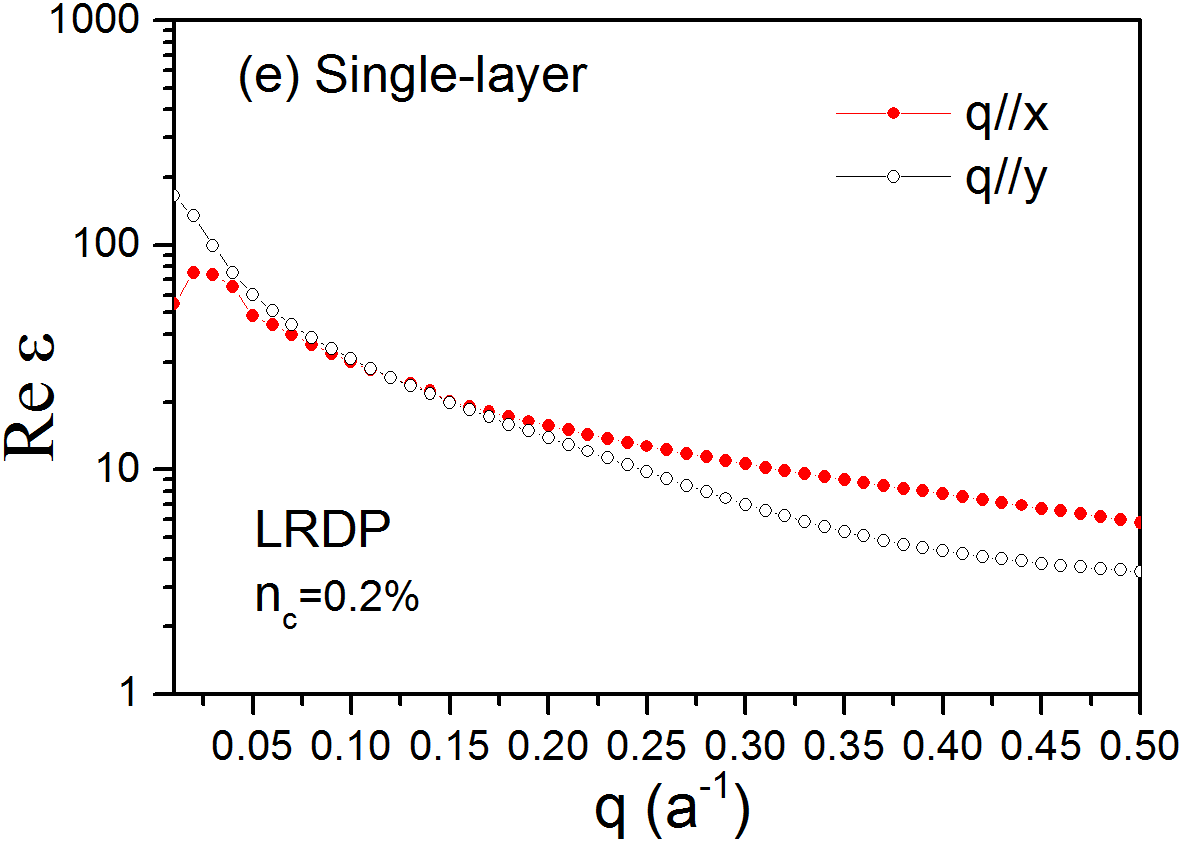} %
\includegraphics[width=4.25cm]{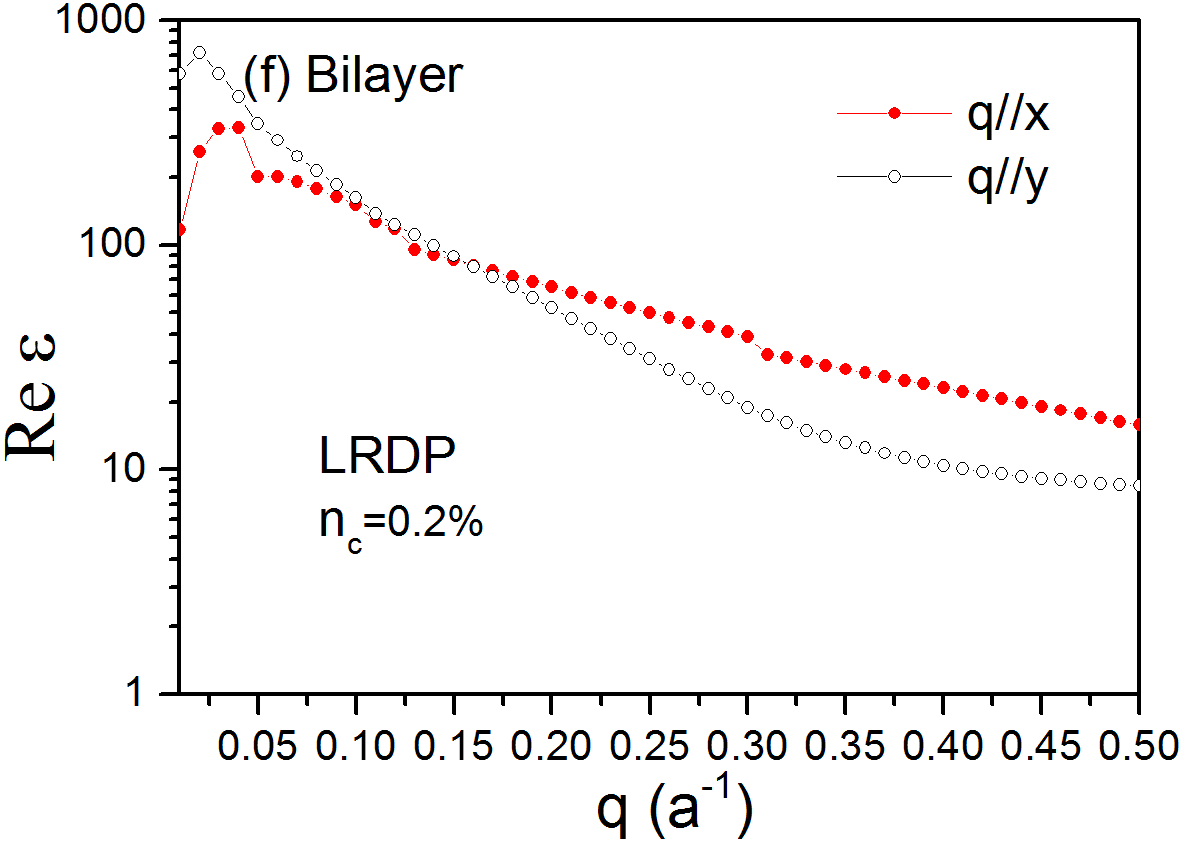}
\end{center}
\caption{Static dielectric function of single- (left) and bilayer (right)
BP. The chemical potential is set to be $\protect\mu =0.4$~eV for
single-layer and $0.73$~eV for bilayer. The temperature is fixed as the room
tempertuare $T=300$~K. Plots (a) and (b) correspond to pristine BP, (c) and
(d) correspond to samples with resonant scatterers originated from a
concentration of $n_x=0.2\%$ vacant atoms, and (e) and (f) correspond to
samples with long range disorder potential with a concentration of $%
n_c=0.2\% $ Gaussian centers.}
\label{Fig:StaticDielectric}
\end{figure}

Electron-electron interactions are considered within the RPA. The
two-dimensional dielectric function for single-layer phosphorene is
calculated as 
\begin{equation}
\mathbf{\varepsilon }_{1L}\left( \mathbf{q},\omega \right) =\mathbf{1}%
-V\left( q\right) \Pi \left( \mathbf{q},\omega \right)  \label{Eq:Epsilon1L}
\end{equation}%
where 
\begin{equation}
V\left( q\right) =\frac{2\pi e^{2}}{\kappa q}  \label{Eq:Coulomb}
\end{equation}%
is the Fourier component of the Coulomb interaction in two dimensions, in
terms of the background dielectric constant $\kappa $. For bilayer BP, the
Coulomb interaction has a matrix form 
\begin{equation*}
{\tilde{\mathbf{V}}}(q)=\left( 
\begin{array}{cc}
V_{\mathrm{intra}}(q) & V_{\mathrm{inter}}(q) \\ 
V_{\mathrm{inter}}(q) & V_{\mathrm{intra}}(q)%
\end{array}%
\right) ,
\end{equation*}%
where $V_{\mathrm{intra}}(q)=V(q)$ as given by Eq. (\ref{Eq:Coulomb}), and
the interlayer interaction between electrons in different BP layers is $V_{%
\mathrm{inter}}=F(q)V(q)$ where $F(q)=e^{-qd}$ and $d=0.5455$~nm is the
interlayer distance. From this we can express the dielectric tensor for
bilayer BP, including the intralayer and interlayer contributions, as%
\begin{eqnarray}
&&\mathbf{\varepsilon }_{2L}\left( \mathbf{q},\omega \right)  \notag \\
&=&\left( 
\begin{array}{cc}
\mathbf{\varepsilon }_{1L}\left( \mathbf{q},\omega \right) & -V\left(
q\right) F\left( q\right) \Pi \left( \mathbf{q},\omega \right) \\ 
-V\left( q\right) F\left( q\right) \Pi \left( \mathbf{q},\omega \right) & 
\mathbf{\varepsilon }_{1L}\left( \mathbf{q},\omega \right)%
\end{array}%
\right) .  \notag \\
&&
\end{eqnarray}

In the next sections we use the polarization and dielectric functions
defined here to study the effect of disorder on the screening properties and
the collective plasmon modes of BP. To simplify the notation, we use $%
\mathbf{\ \varepsilon }\left( \mathbf{q},\omega \right) $ to represent $%
\mathbf{\varepsilon }_{1L}\left( \mathbf{q},\omega \right) $ for
single-layer and the determinant of the dielectric tensor $\left\vert 
\mathbf{\varepsilon }_{2L}\left( \mathbf{q},\omega \right) \right\vert $ for
bilayer.

\section{Dielectric screening}\label{Sec:Screening}

We start by studying the effect of disorder on the screening properties. In
Fig.~\ref{Fig:StaticDielectric}, we plot the static dielectric function $%
\mathbf{\varepsilon }\left( \mathbf{q},0\right) $ of single- and bilayer BP.
There is a clear dependence of the static dielectric function with the
direction of momentum, showing a different behavior for $\mathbf{q}$ along
the zigzag (X) or armchair (Y) directions. For the pristine case, our
numerical results for the momentum dependence of the static dielectric
function of single-layer phosphorene, Fig. \ref{Fig:StaticDielectric}(a),
agree with the analytic low energy model presented in Ref. %
\onlinecite{Low2014p}. The present calculation within the TBPM method allows
us to extend the calculation of Ref. \onlinecite{Low2014p} to larger
wave-vectors and to consider realistic impurities in the sample. We notice that the size of the lattices used in our simulations determines the rage of wave-vectors for our numerical results, which is $ q \geq0.01a^{-1}$. We do not perform calculations with momentum transfer smaller than $q=0.01a^{-1}$~since, in order to preserve the
numerical accuracy for $q\rightarrow 0$, an extreamly large sample size  $%
N\rightarrow \infty $ is required.

In both single- and bilayer BP, point-like resonant scatterers lead to the
creation of midgap states inside the gap.\cite{Yuan2015BP} These states are
quasi-localized around the center of the impurity. Therefore, the typical
divergence of the dielectric function at $q\rightarrow 0$, characteristic of
metallic behavior, is reduced for BP crystals with this kind of disorder, as
it can be seen in Fig. \ref{Fig:StaticDielectric}(c)-(d). A similar result
for the dielectric function is obtained for LRDP which can be associated to
the existence of electron-hole puddles in the sample. A drop of the
dielectric function at small $q$ appears also in disordered graphene with
chemical absorbers. \cite{Yuan2012} The effect of disorder on the static
dielectric function is negligible at short wave-lengths (large $q$ vectors).

\section{Energy Loss Function and Plasmons}\label{Sec:Plasmons}

\begin{figure}[t]
\begin{center}
\mbox{
\includegraphics[width=4.25cm]{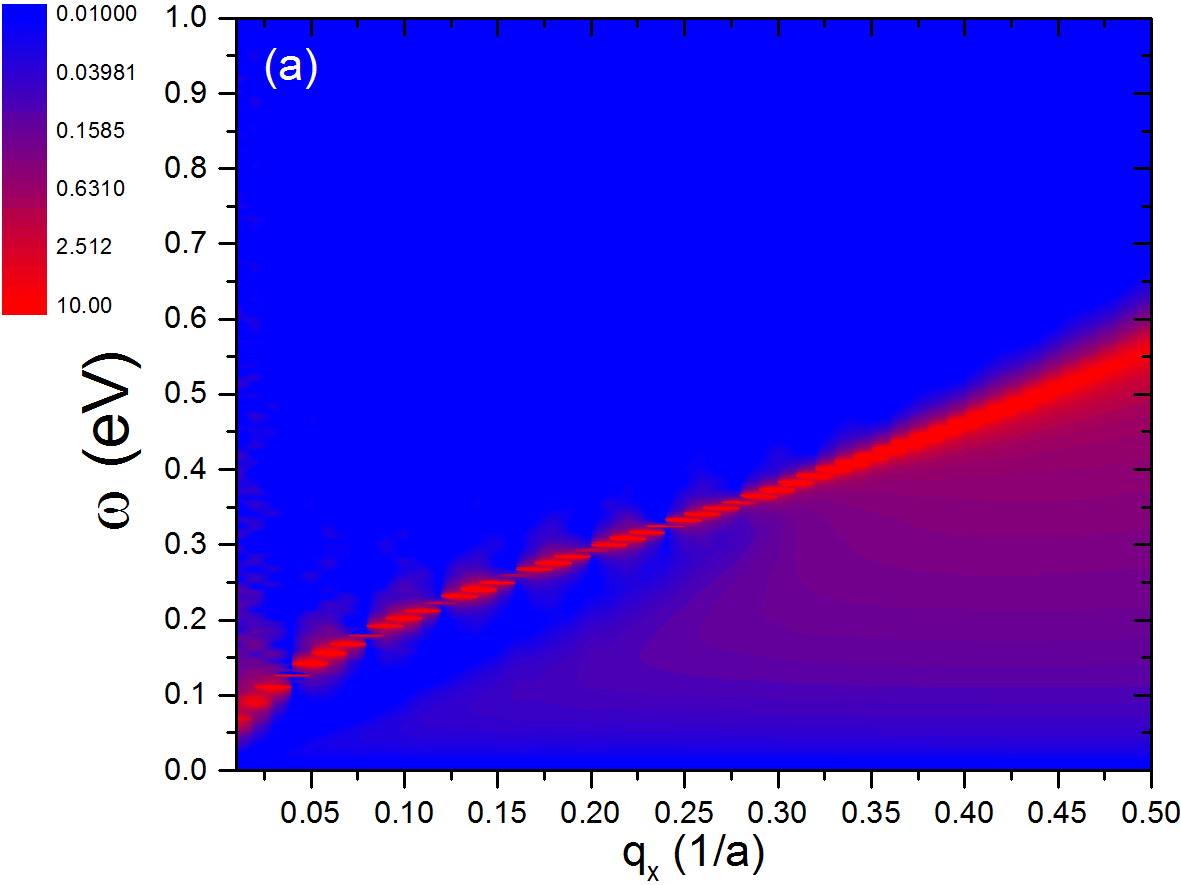}
\includegraphics[width=4.25cm]{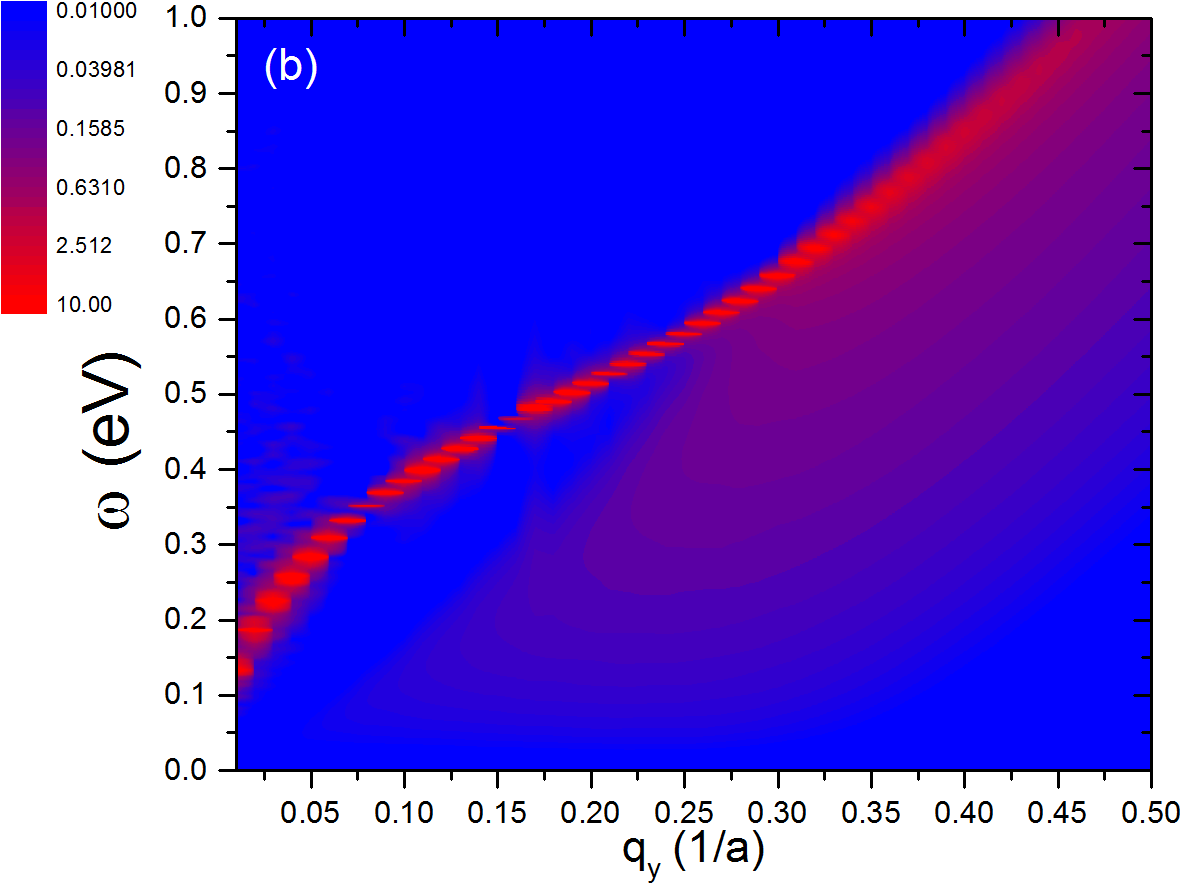}
} 
\mbox{
\includegraphics[width=4.25cm]{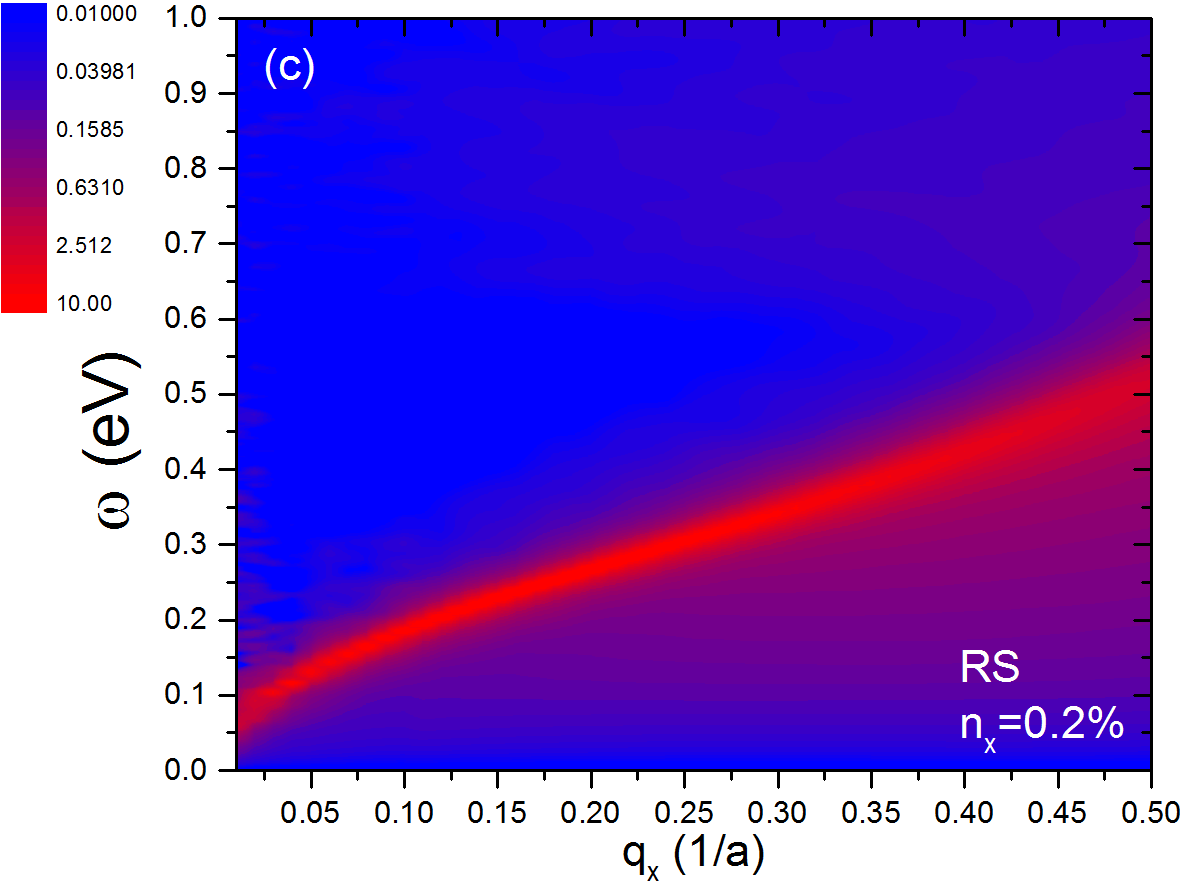}
\includegraphics[width=4.25cm]{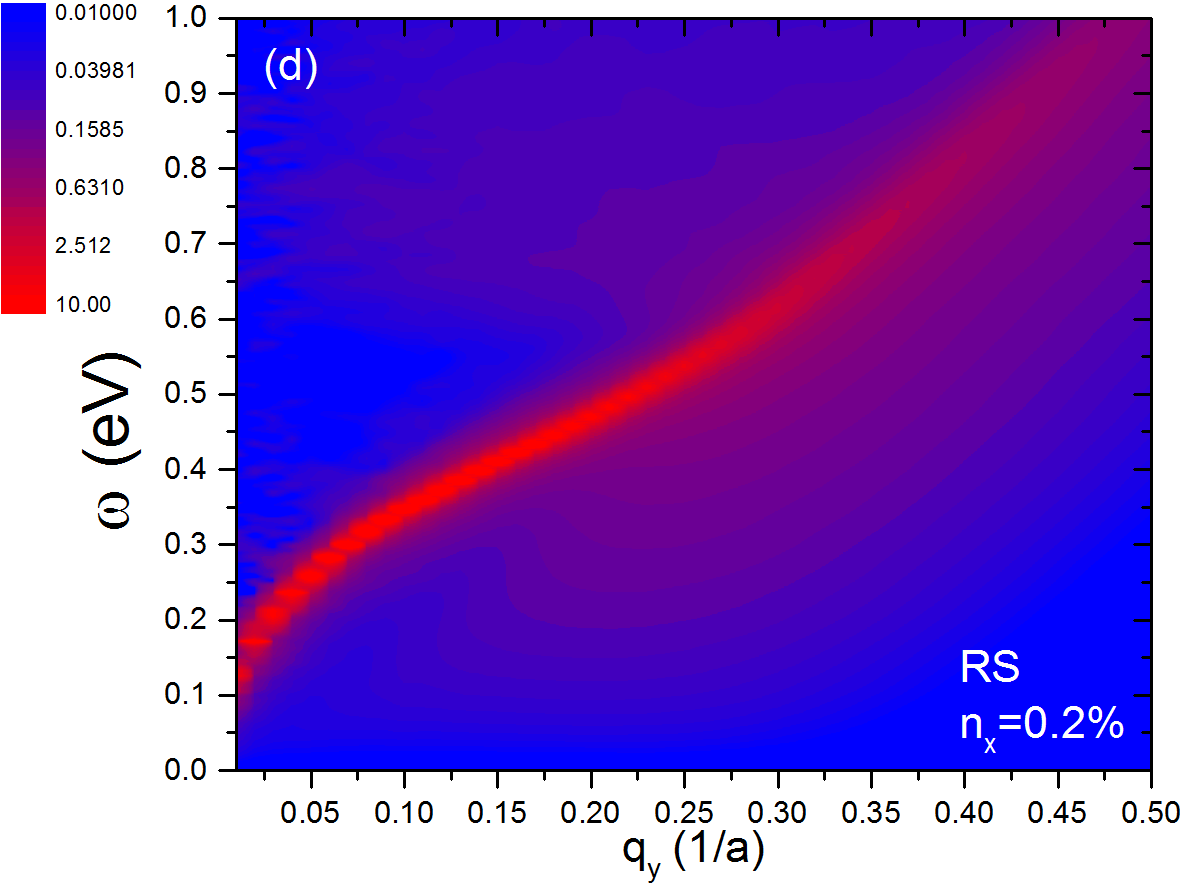}
} 
\mbox{
\includegraphics[width=4.25cm]{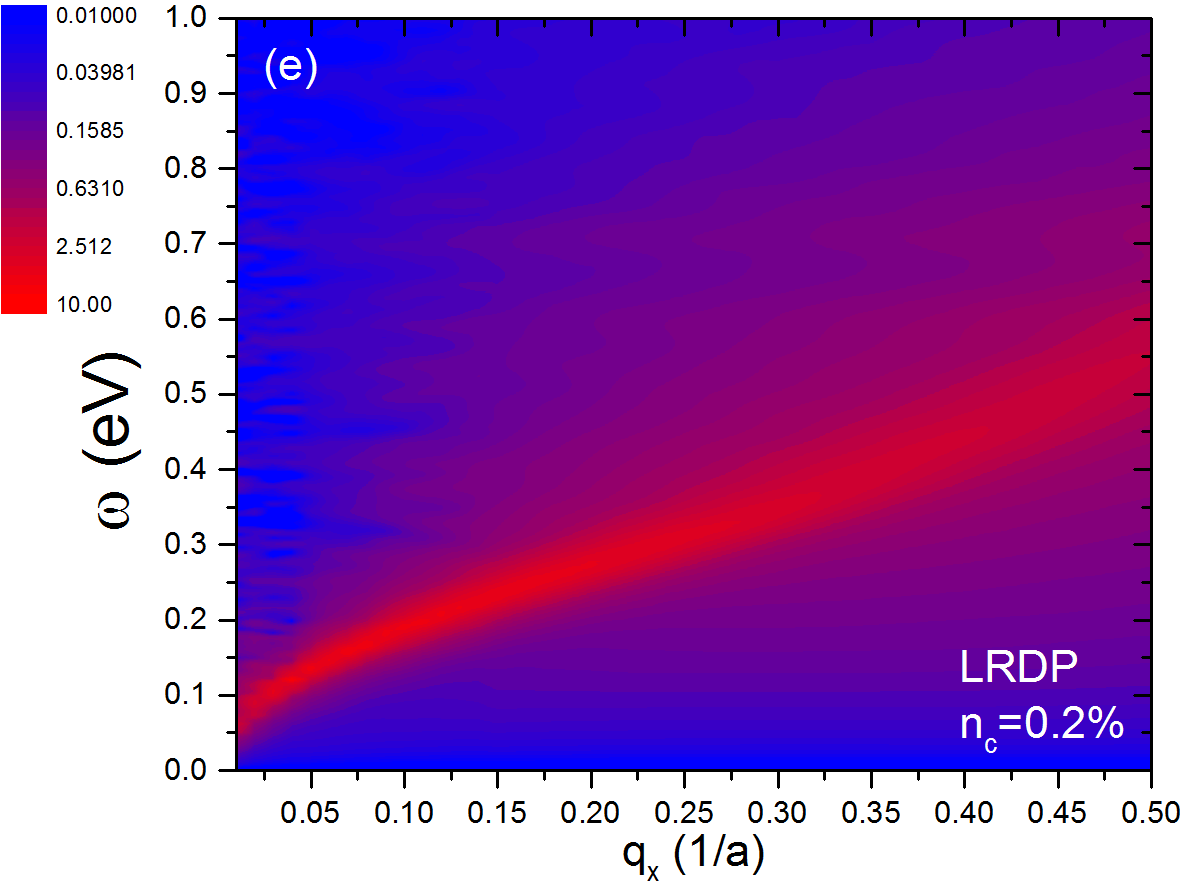}
\includegraphics[width=4.25cm]{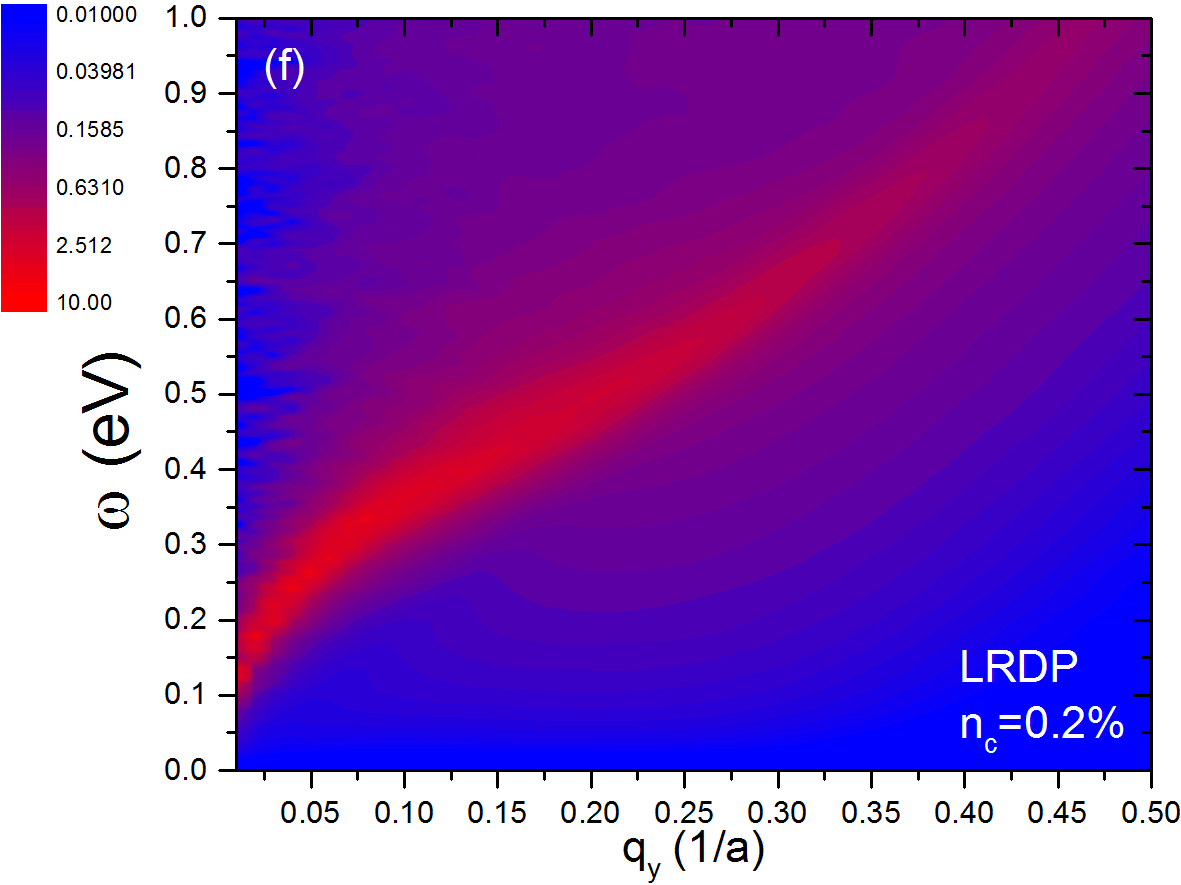}
}
\end{center}
\caption{Energy loss functions of single-layer BP along zigzag (left) and
armchair (right) directions. The simulations are done for: (a,b) pristine
BP, (c,d) samples with resonant scatterers, and (e,f) samples with
long-range disorder potentials. All the results are calculated at $T=300$~K. 
We notice that the apparent discretization of the spectrum in some of the plots is an artifact due to finite size limitations in our calculations.}
\label{Fig:EnergyLossSingle}
\end{figure}

\begin{figure}[tb]
\begin{center}
\mbox{
\includegraphics[width=4.25cm]{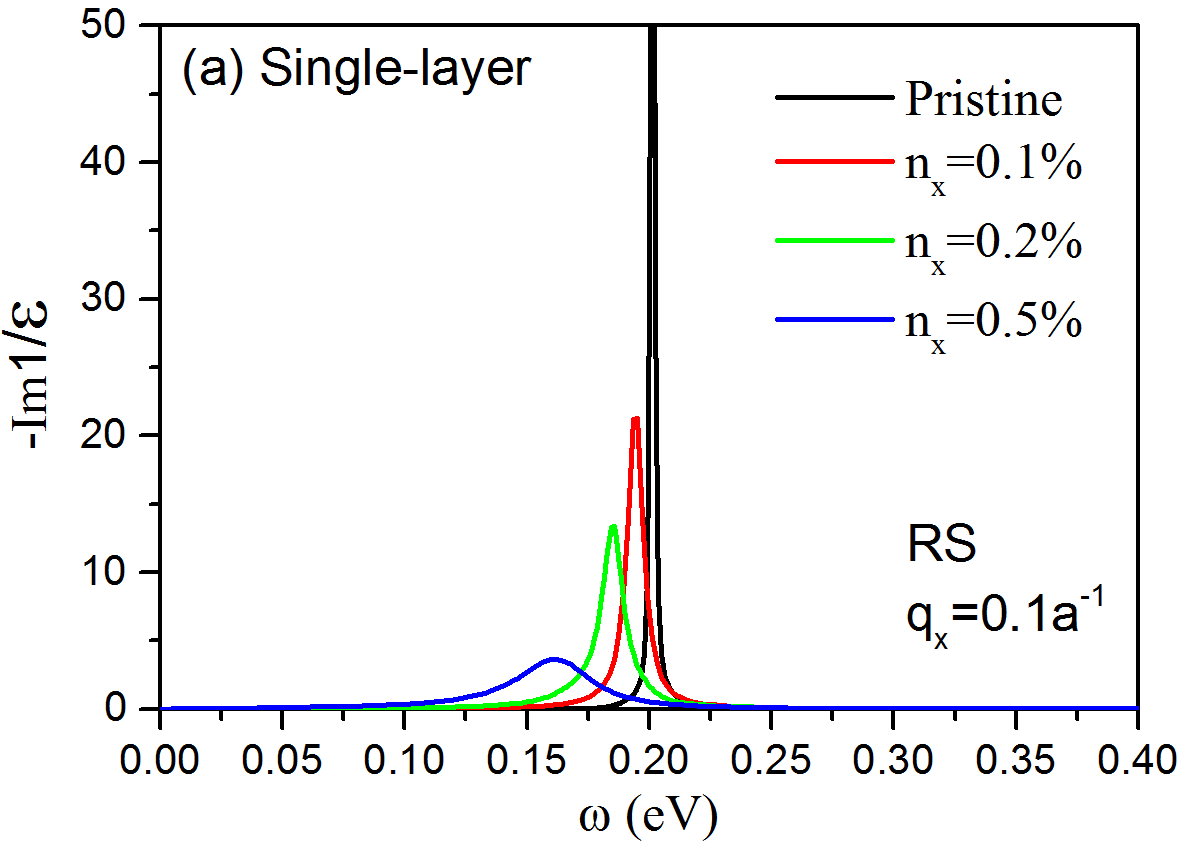}
\includegraphics[width=4.25cm]{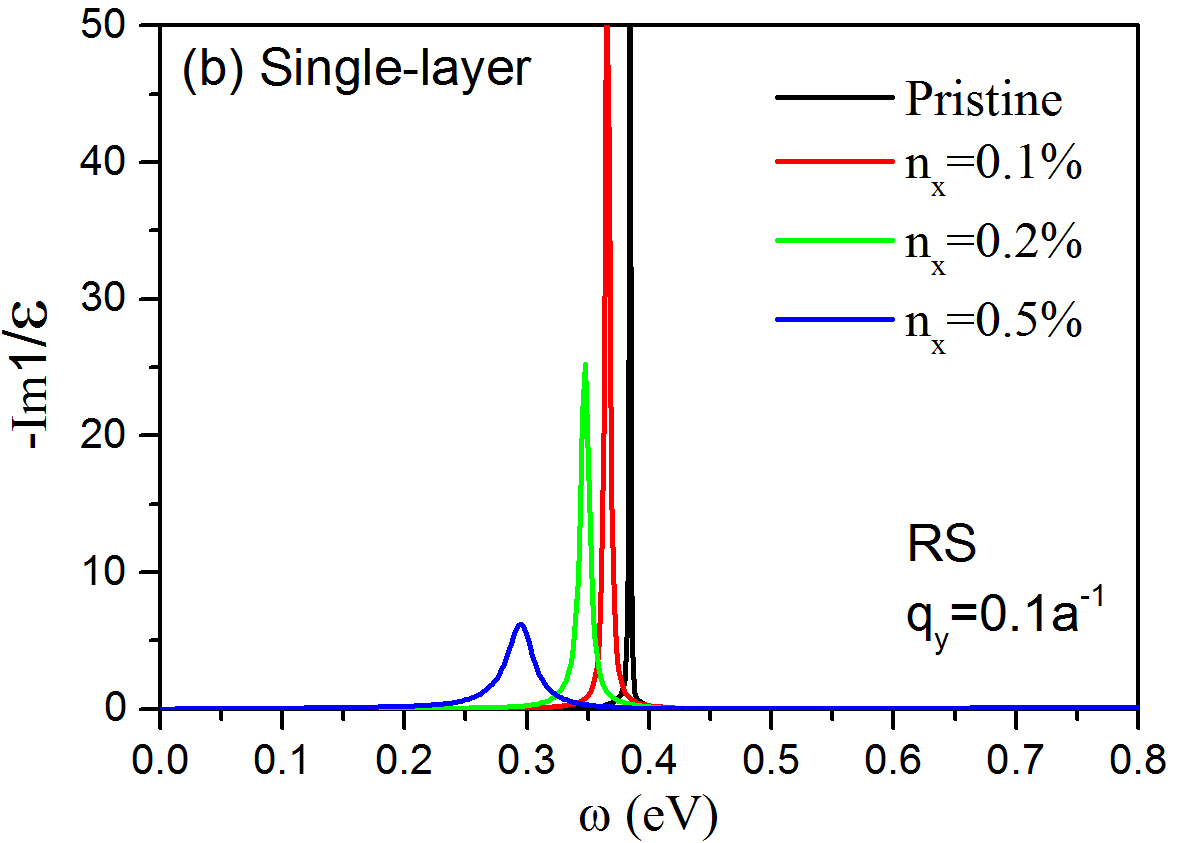}
} 
\mbox{
\includegraphics[width=4.25cm]{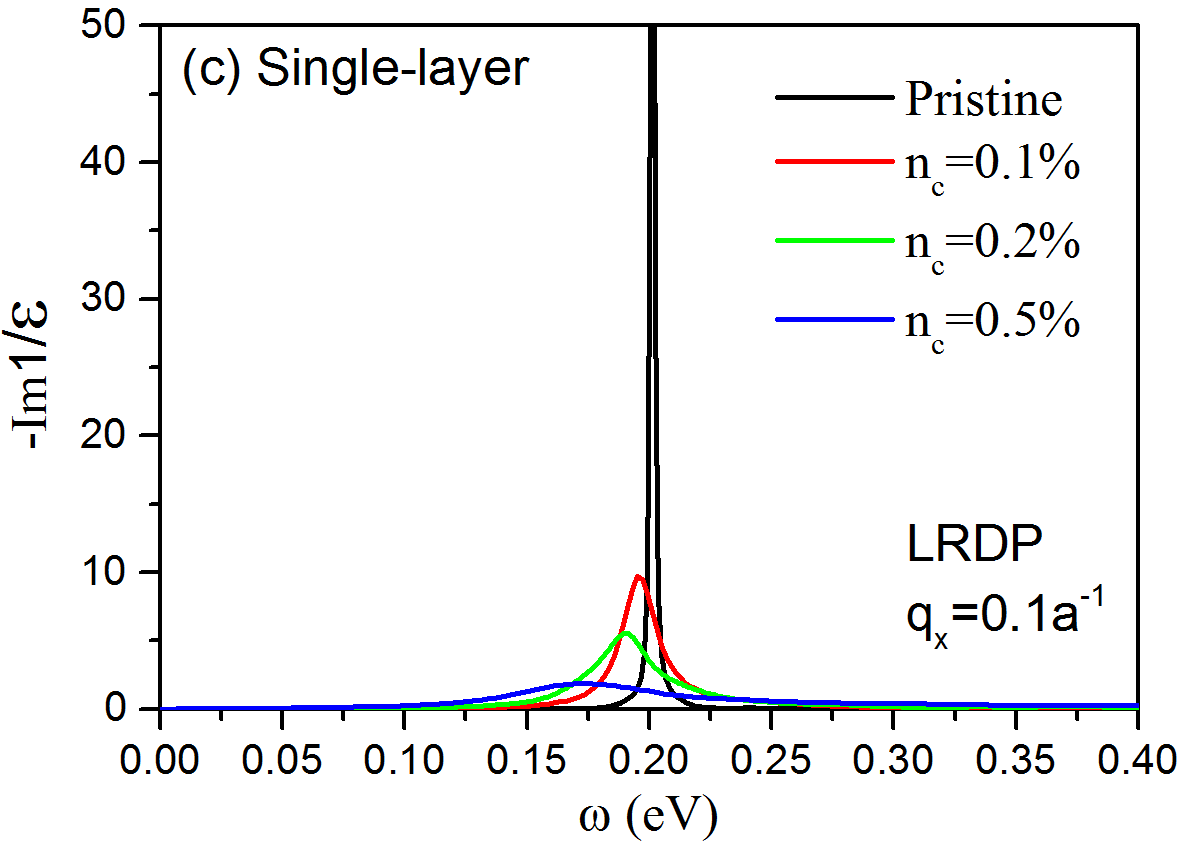}
\includegraphics[width=4.25cm]{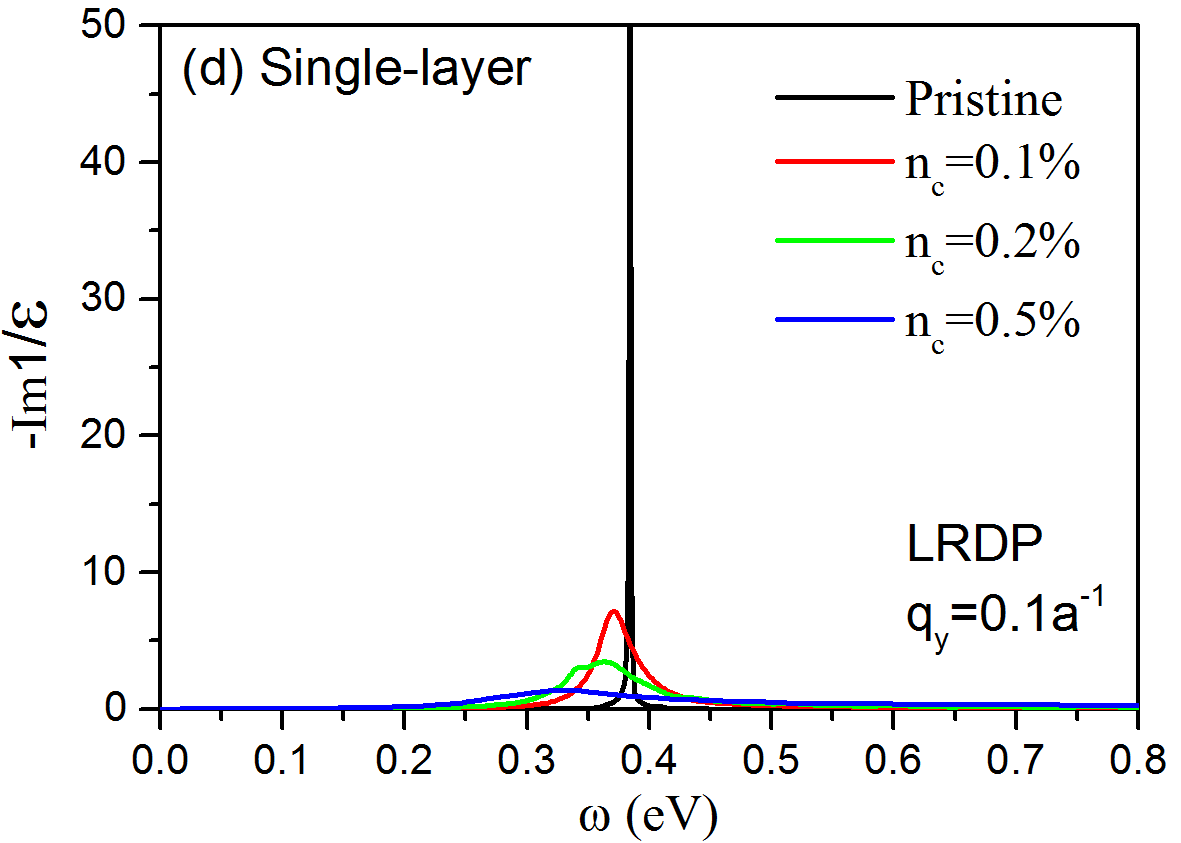}
}
\end{center}
\caption{Energy loss functions of single-layer BP along zigzag (left) and
armchair (right) directions. Panels (a) and (b) compare the loss function of
pristine single-layer BP with that of samples with different concentrations
of point defects. Panels (c) and (d) correspond to LRDP. All the results are
calculated at $T=300$~K.}
\label{Fig:EnergyLossSingle-q}
\end{figure}

In this section we study the collective modes of the system, which are
defined by the zeroes of the dielectric function $\mathbf{\varepsilon }%
\left( \mathbf{q},\omega \right)$ for single-layer BP (or zeroes of the
dielectric tensor determinant for bilayer BP). The dispersion relation of
the plasmon modes is defined from $\mathrm{Re}~\varepsilon (\mathbf{q}%
,\omega _{pl})=0, $ which leads to poles in the energy loss function $-\Im
1/\varepsilon (q,\omega )$, that can be measured by means of electron energy
loss spectroscopy (EELS).\cite{Politano2014}

\subsection{Single-layer BP}

For single-layer BP, our numerical results for the loss function are plotted
in Fig. \ref{Fig:EnergyLossSingle} and show anisotropic plasmon modes along
the zigzag and armchair directions, both following a low energy $\sqrt{q}$
dependence at small momentum transfer. This is due to the paraboloidal band
dispersion of BP at low energies, and it is consistent with the results for
the excitation spectrum of BP within the $\mathbf{k\cdot p}$ approximation.%
\cite{Low2014p} The present method allows us to study the spectrum at higher
energies and wave-vectors, including also the effect of disorder in the
sample.

In the presence of disorder, there is a broadening and red shift of the
plasmon modes, as it can be seen in Fig. \ref{Fig:EnergyLossSingle}(c)-(d)
for single-layer samples with resonant scatterers, and in Fig. \ref%
{Fig:EnergyLossSingle}(e)-(f) for long range disorder potential. For a fixed
wave-vector, the effect of disorder is better seen by the loss function
shown in Fig. \ref{Fig:EnergyLossSingle-q} for $q=0.1a^{-1}$ and different
concentrations of disorder. Notice that the presence of LRPD induces a
stronger damping of the plasmon, as compared to point defects. This is seen
by comparing the broadening of the peaks of the top and bottom panels of
Fig. \ref{Fig:EnergyLossSingle-q}. Since plasmons are collective electronic
oscillations induced by long range Coulomb interaction, it is expected that
long range disorder affects more the coherence of these modes than local
scatterers as vacancies, which are relevant at short length scales.

Fig. \ref{Fig:EnergyLossSingle-q} also shows that, for a given wavelength,
the two kinds of disorder considered in our simulations lead to a shift of
the plasmon resonance towards lower energies. This effect can be understood
from a perturbative point of view, by using the disordered averaged response
function introduced by Mermin in the context of a 2DEG with a parabolic band
dispersion,\cite{Mermin70} and which has been recently used to study the
effect of weak disorder on the plasmon dispersion of spin-polarized graphene.%
\cite{AV15} In the weak disorder limit, and assuming for simplicity an 
\textit{isotropic} parabolic band dispersion, the disordered averaged
response function has the form\cite{Mermin70,GV05} 
\begin{equation}  \label{Eq:Pi0Imp}
\Pi_{imp}(\mathbf{q},\omega)\approx -N(0)\left(1-\frac{\omega}{\sqrt{%
(\omega+i/\tau_{el})^2-v_F^2q^2}-i/\tau_{el}}\right)
\end{equation}
where $\tau_{el}$ is the elastic life-time of the momentum eigenstates of
the disordered system, $N(0)=gm_b/2\pi$ is the density of states (per unit
area) at the Fermi energy, where $g=2$ is the spin degeneracy, $m_b$ is the
effective mass, and $v_F$ is the Fermi velocity. In the presence of disorder
the dispersion relation of the plasmon will have, even in the low energy and
long wavelength limit, and imaginary part, due to finite damping of the
plasmon. Within this approximation, it is possible to obtain an analytical
expression for the plasmon dispersion by substituting (\ref{Eq:Pi0Imp}) into
(\ref{Eq:Epsilon1L}) and then finding the zeros of the dielectric function,
with the approximate solution 
\begin{equation}  \label{Eq:PlasmonImp}
\omega_{pl}(q)\approx \frac{q+4r_sk_F}{q+8r_sk_F}\left(\sqrt{%
v_F^2q(q+8r_sk_F)-1/\tau_{el}^2}-i/\tau_{el}\right)
\end{equation}
where $r_s=2m_be^2/\kappa k_F$ is the dimensionless interaction parameter of
a 2DEG and $k_F$ is the Fermi wave-vector. Eq. (\ref{Eq:PlasmonImp}) reduces
to the standard plasmon dispersion in the clean ($1/\tau_{el}\rightarrow 0$)
limit. We notice here that the presence of impurities limit the dispersion
of collective plasmon modes for length scales longer than the mean free
path, defining a minimal wave-vector below which the plasmon mode is not
well defined. The existence of such length scale has been discussed, in the
context of graphene, in Ref. \onlinecite{AV15}. For the case of interest
here such infrared cutoff takes the form $q_{min}=-4r_sk_F+\sqrt{%
8r_s^2k_F^2+1/\tau_{el}^2 v_F^2}$, below which plasmon modes do not exist.

The next step in our analysis is to quantify the effect of disorder on
plasmon losses. The physical observability of a damped collective mode
depends on the sharpness of the resonance peak, and it is defined as a
dimensionless inverse quality factor $\mathcal{R}=\gamma /\omega _{pl}$,
where the damping rate $\gamma $ of the mode, which is proportional to $%
\mathrm{Im}~\Pi (\mathbf{q},\omega _{pl})$, is given by 
\begin{equation}
\gamma =\frac{\mathrm{Im}~\Pi (\mathbf{q},\omega _{pl})}{\left. \frac{%
\partial }{\partial \omega }\mathrm{Re}~\Pi (\mathbf{q},\omega )\right\vert
_{\omega =\omega _{pl}}}.  \label{Eq:Damping}
\end{equation}

The plasmon lifetime $\tau $ is calculated via the inverse quality factor as 
$\tau =1/\mathcal{R}\omega _{pl}.$ We have calculated numerically $\mathcal{R%
}$ ($\gamma $) and $\tau $, and the results are shown in Fig. \ref%
{Fig:Lifetime}. We first notice that, for a given concentration of
impurities and for a given wavelength, the plasmon rate is anisotropic. In
fact, our results show that plasmons dispersing in the $y$-direction
(armchair) are more efficiently damped than plasmons dispersing in the $x$
(zigzag) direction, which have a larger lifetime. Second, as discussed
above, we find that short range resonant scatterers like point defects cause
less losses in the plasmon coherence than LRDP.

\begin{figure}[tb]
\begin{center}
\mbox{
\includegraphics[width=4.25cm]{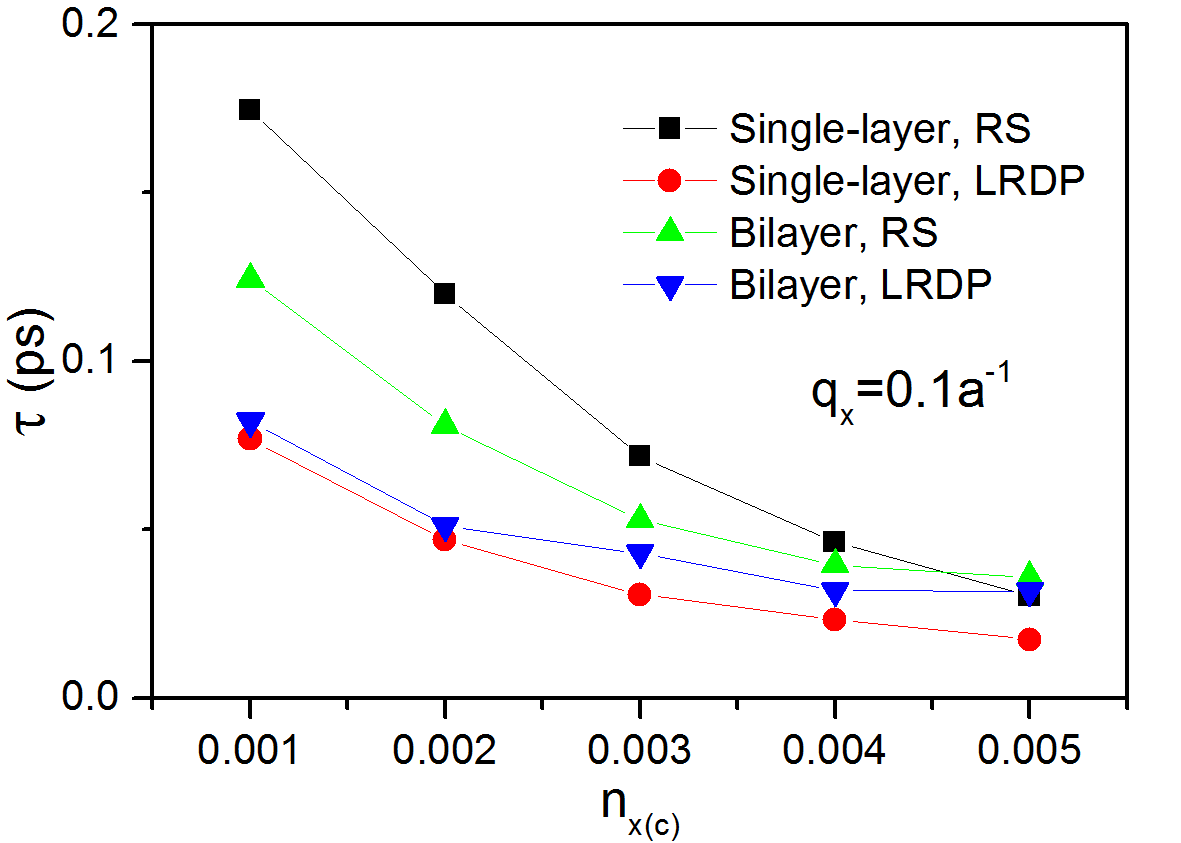}
\includegraphics[width=4.25cm]{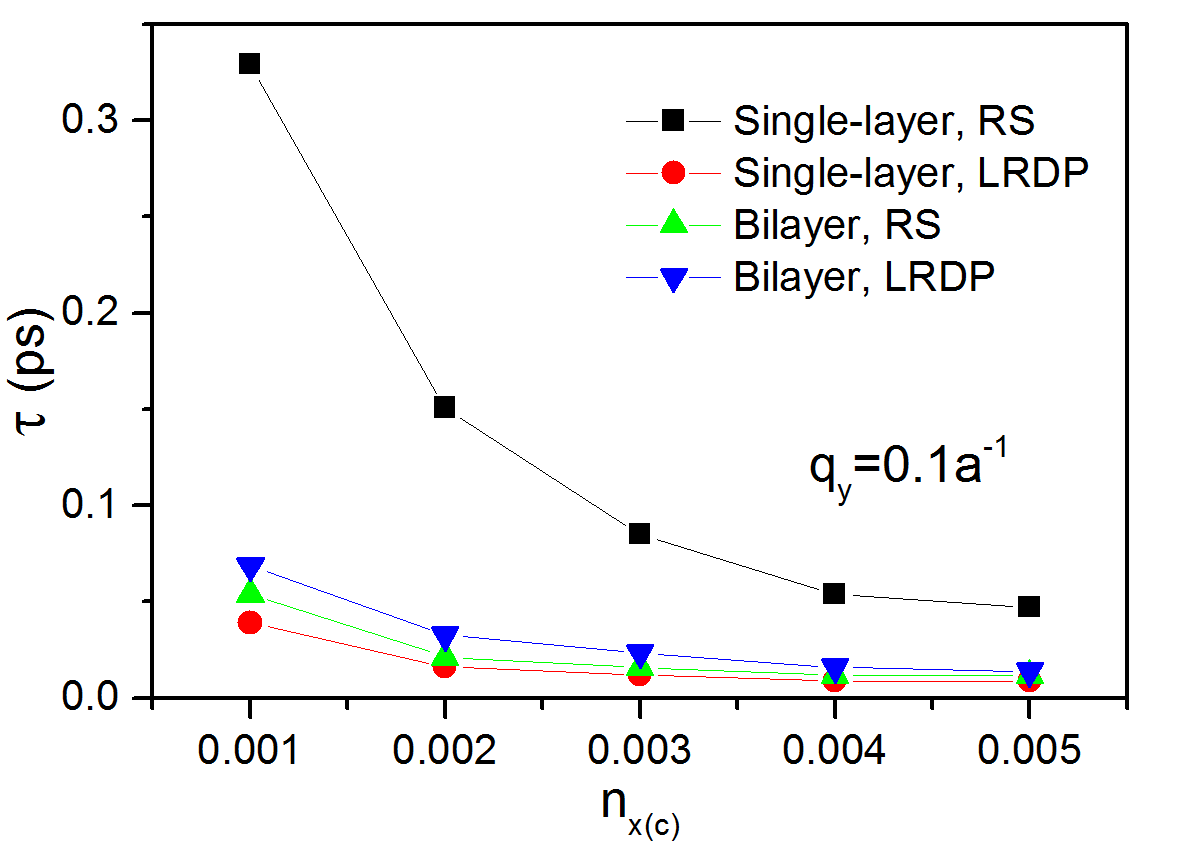}
} 
\mbox{
\includegraphics[width=4.25cm]{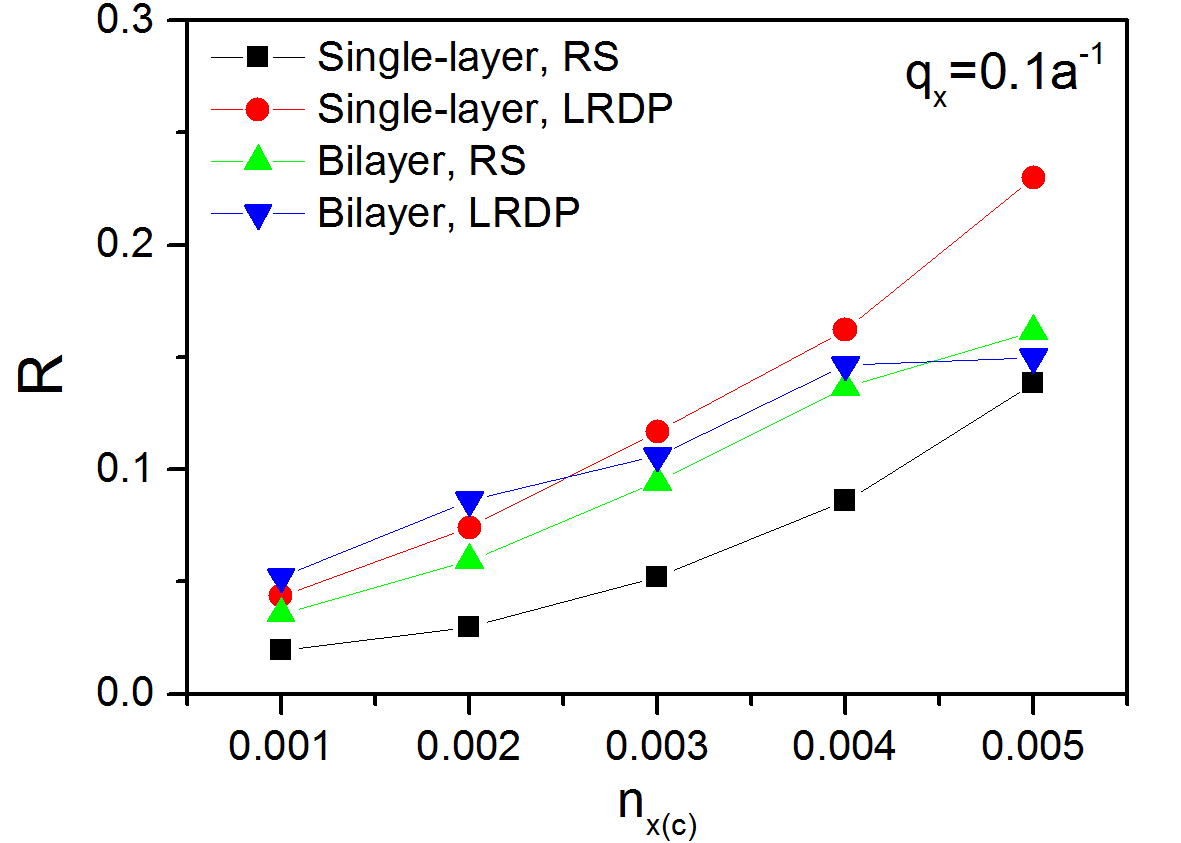}
\includegraphics[width=4.25cm]{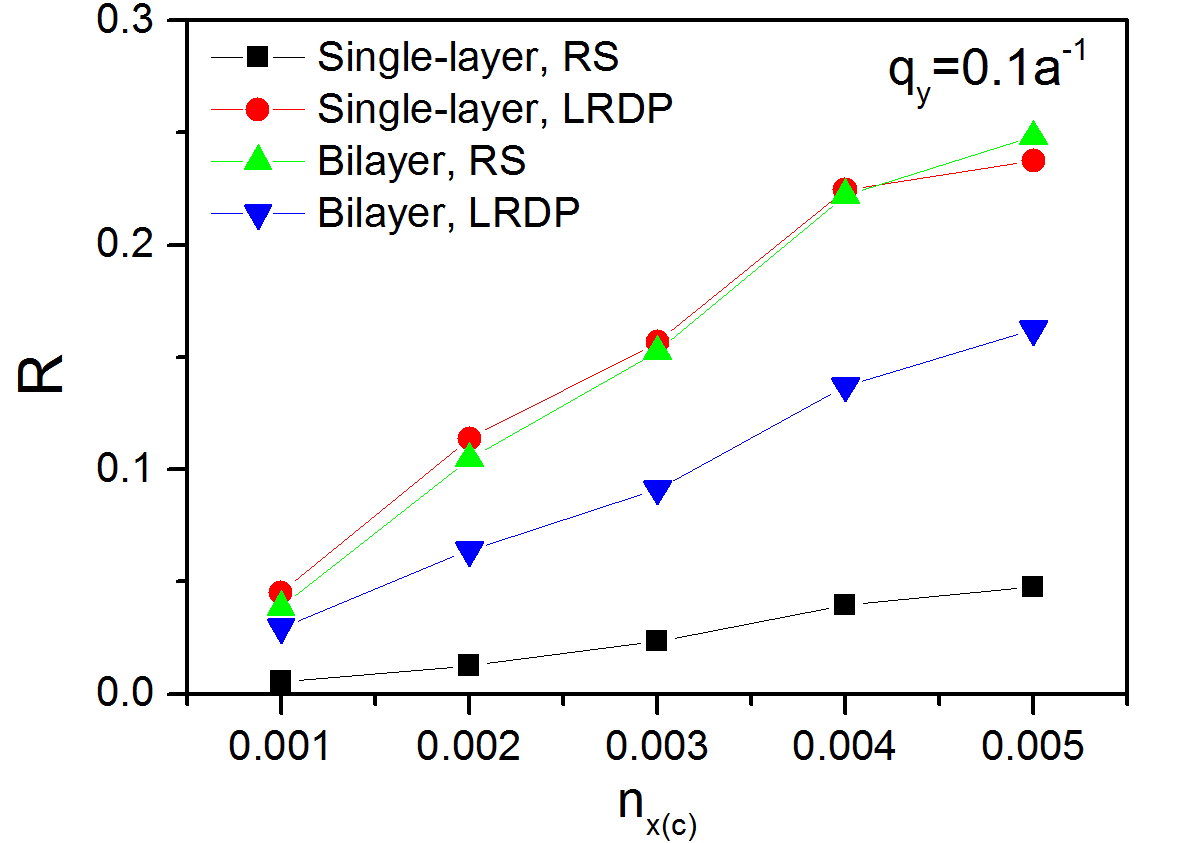}
}
\end{center}
\caption{Effect of disorder on plasmon losses. Panels (a) and (b) show the
plasmon life-time, and panels (c) and (d) show the corresponding damping
rates (inverse quality factor). The results for bilayer BP samples
correspond to the in-phase $\sim \protect\sqrt{q}$ plasmon mode (see text).}
\label{Fig:Lifetime}
\end{figure}

\begin{figure}[t]
\begin{center}
\mbox{
\includegraphics[width=4.25cm]{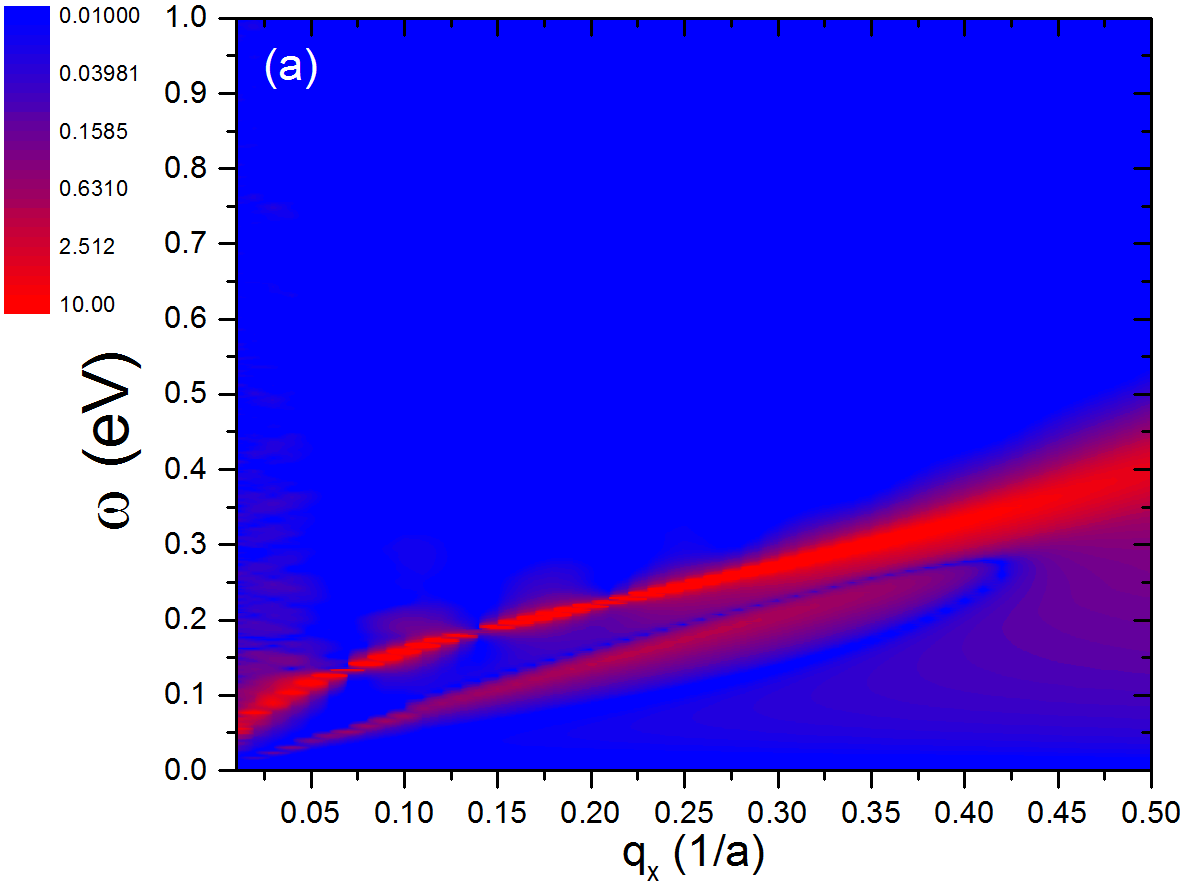}
\includegraphics[width=4.25cm]{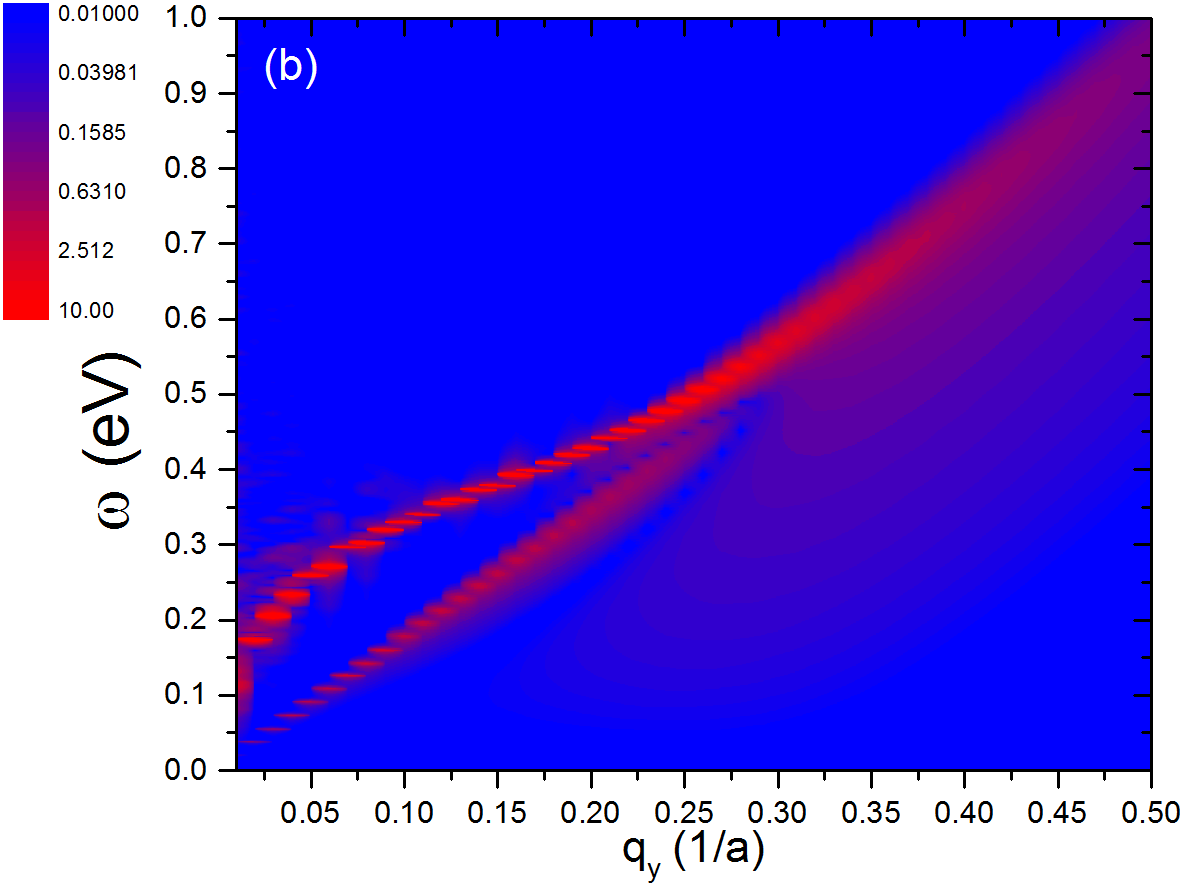}
} 
\mbox{
\includegraphics[width=4.25cm]{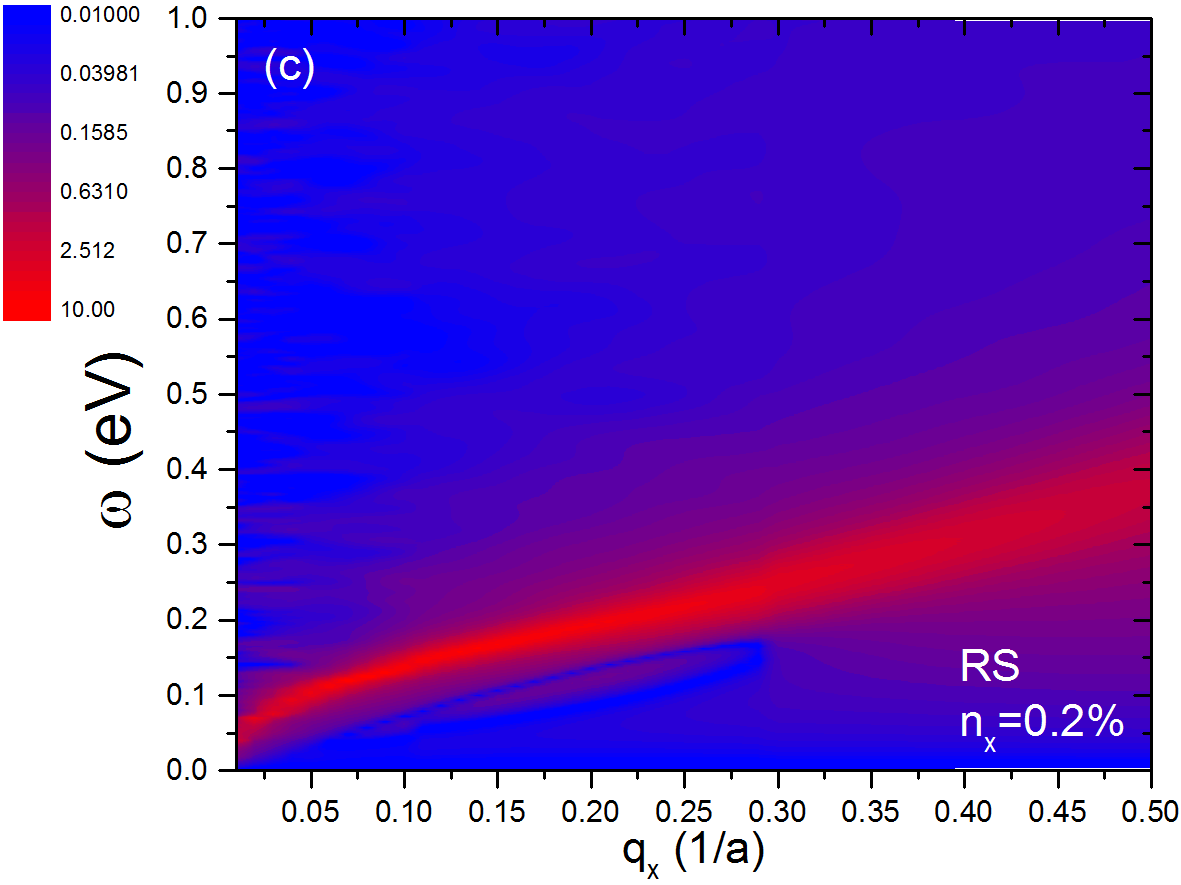}
\includegraphics[width=4.25cm]{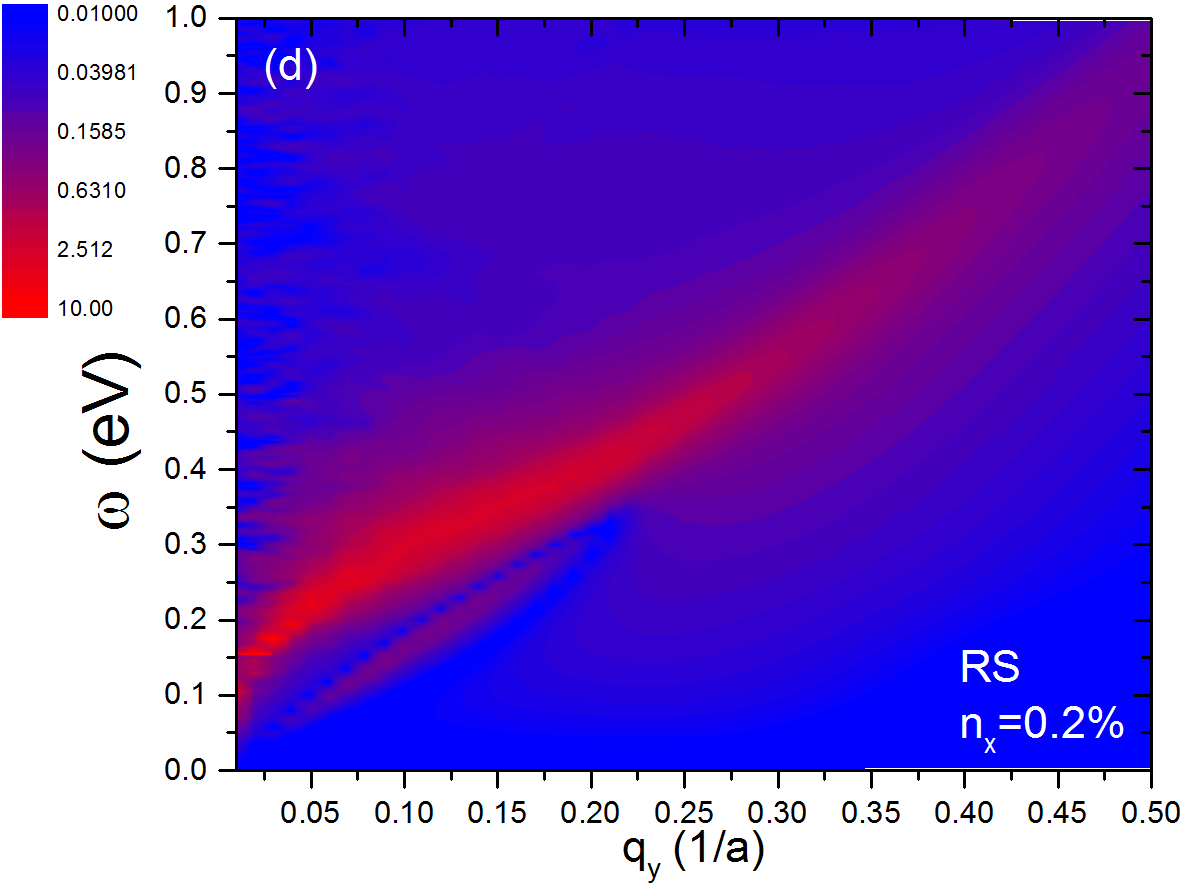}
} 
\mbox{
\includegraphics[width=4.25cm]{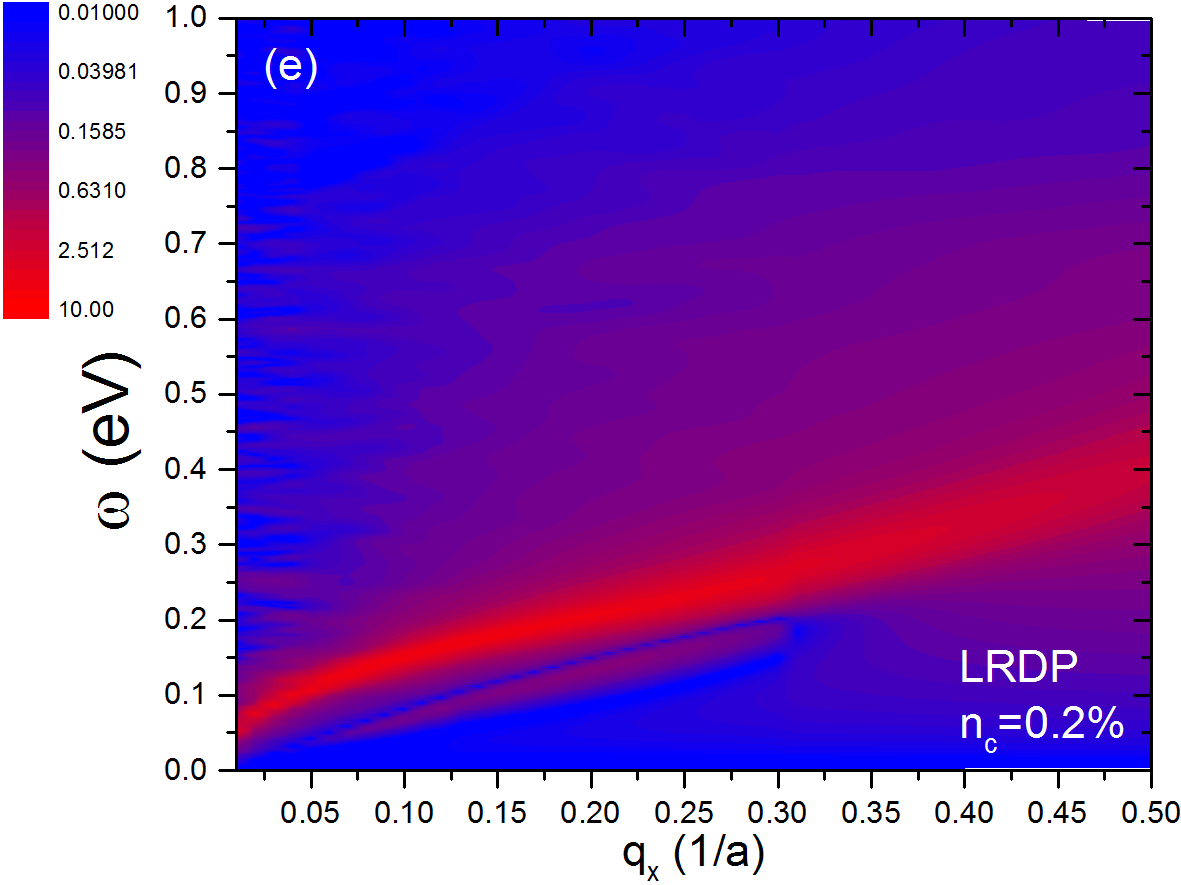}
\includegraphics[width=4.25cm]{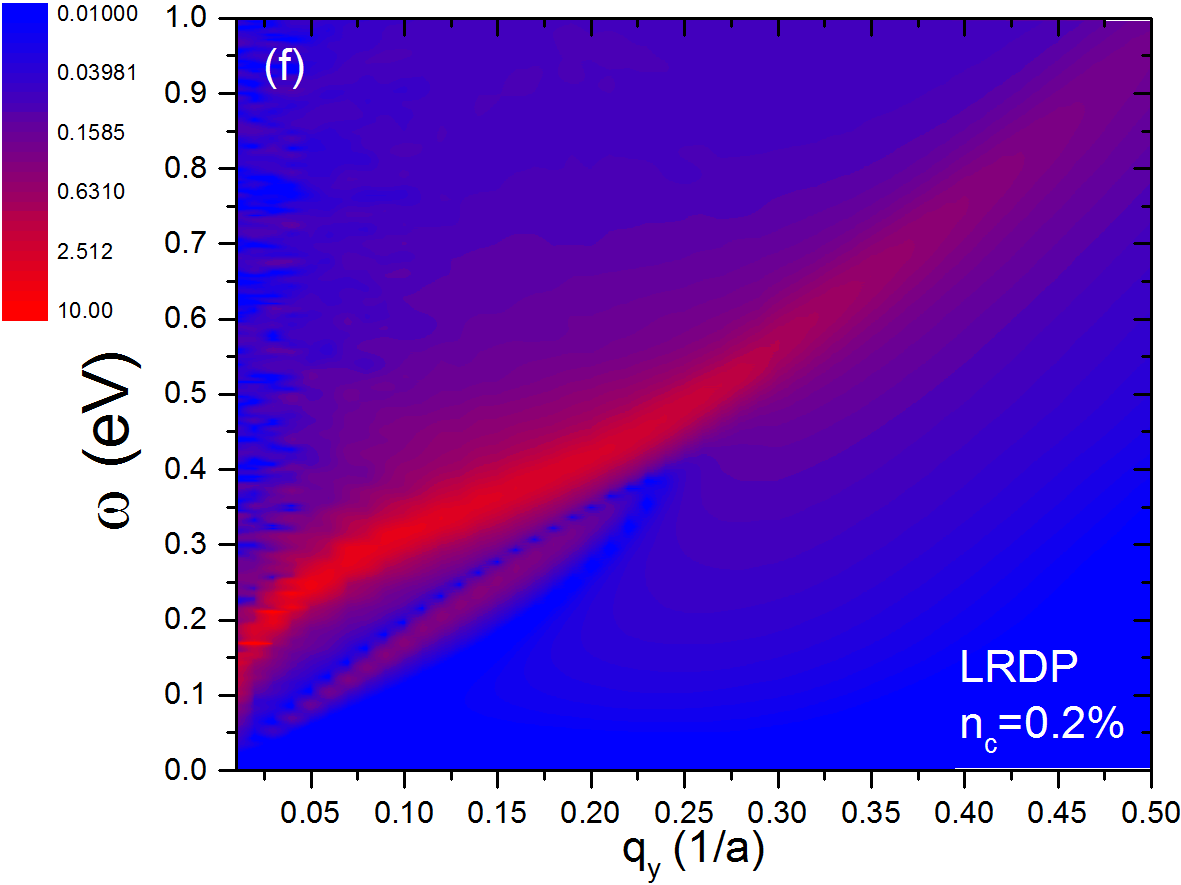}
}
\end{center}
\caption{Energy loss functions of bilayer BP along zigzag (left) and
armchair (right) directions. The simulations are done for: (a,b) pristine
bilayer BP, (c,d) samples with resonant scatterers, and (e,f) samples with
long-range disorder potentials. All the results are calculated at $T=300$~K. 
}
\label{Fig:EnergyLossbilayer}
\end{figure}

\subsection{Bilayer BP}

\begin{figure}[tb]
\begin{center}
\mbox{
\includegraphics[width=4.25cm]{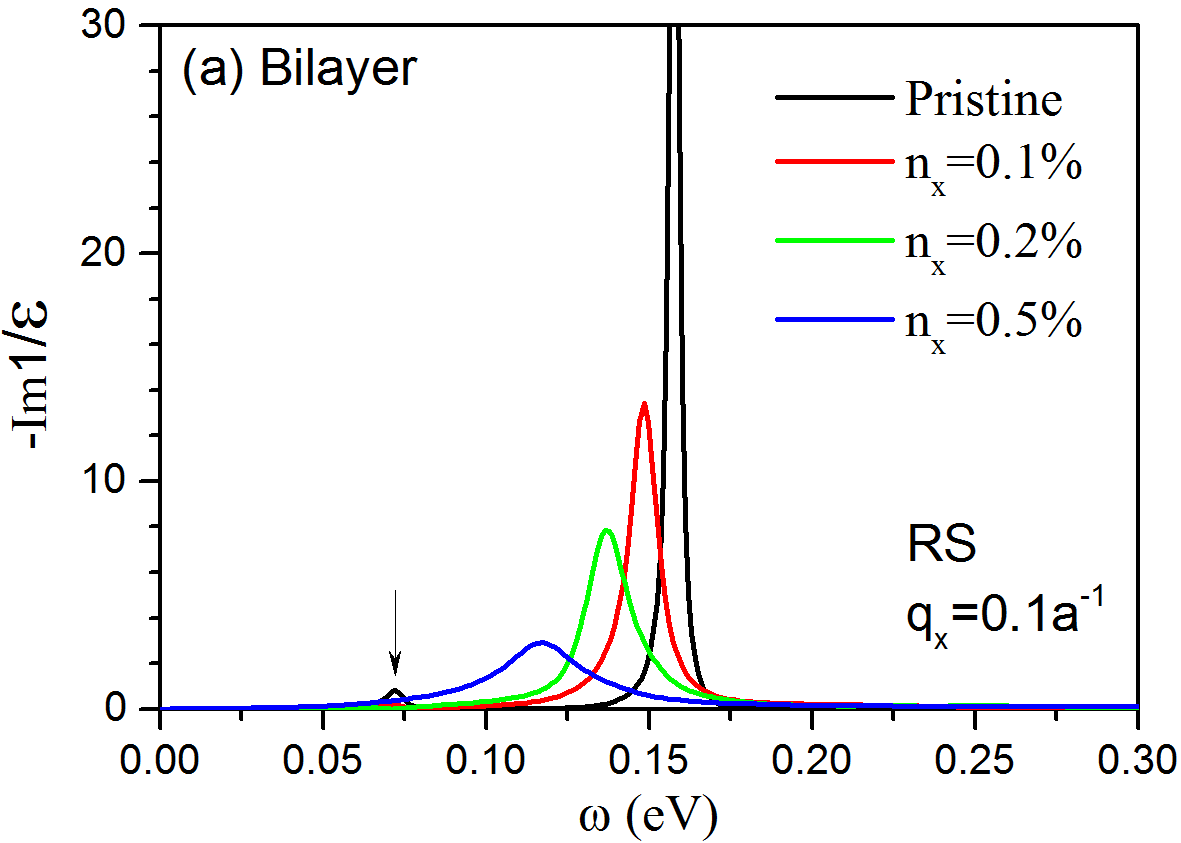}
\includegraphics[width=4.25cm]{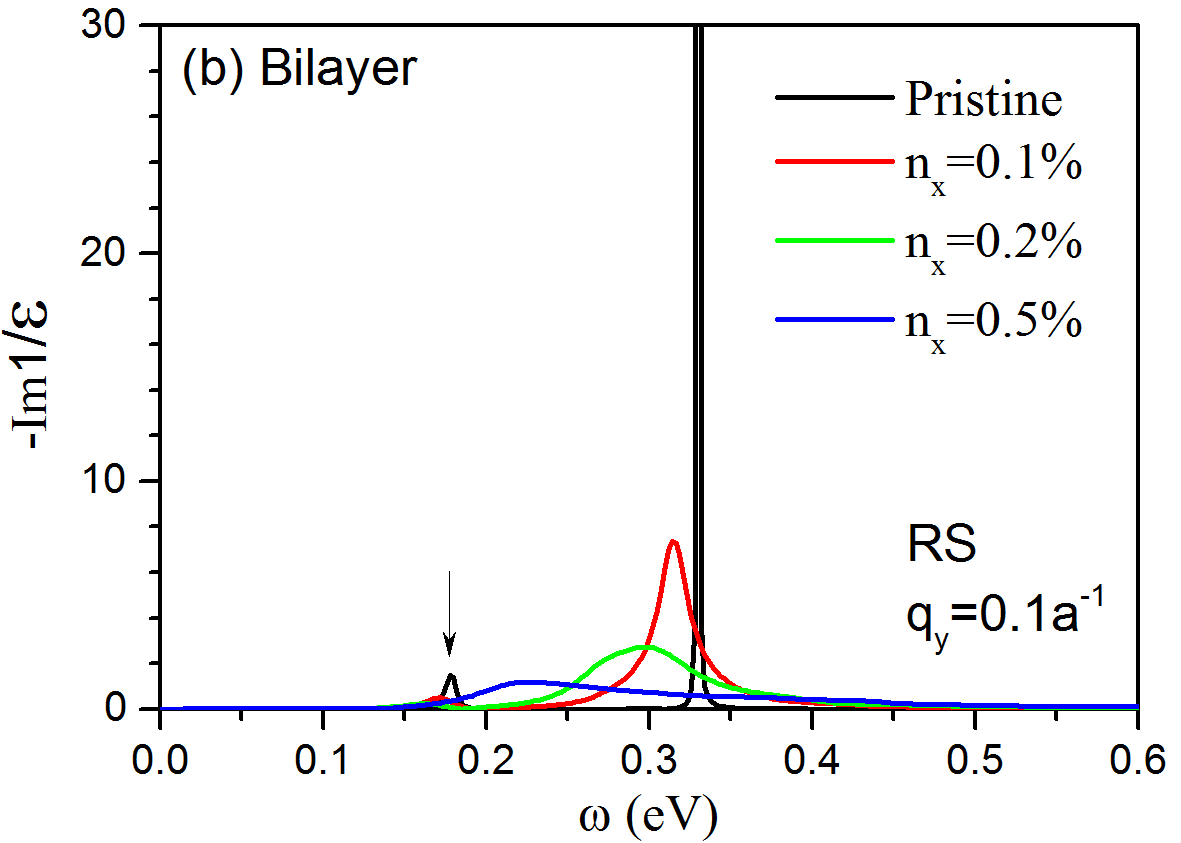}
} 
\mbox{
\includegraphics[width=4.25cm]{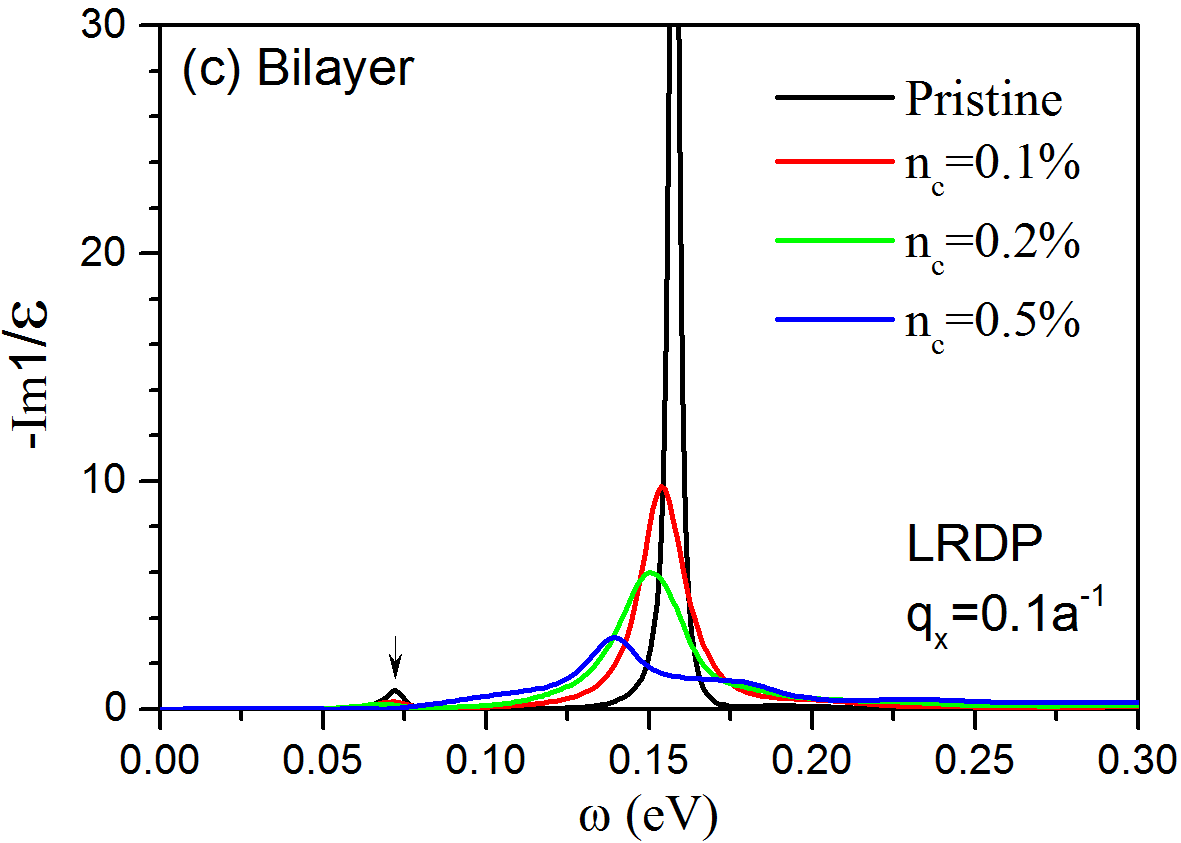}
\includegraphics[width=4.25cm]{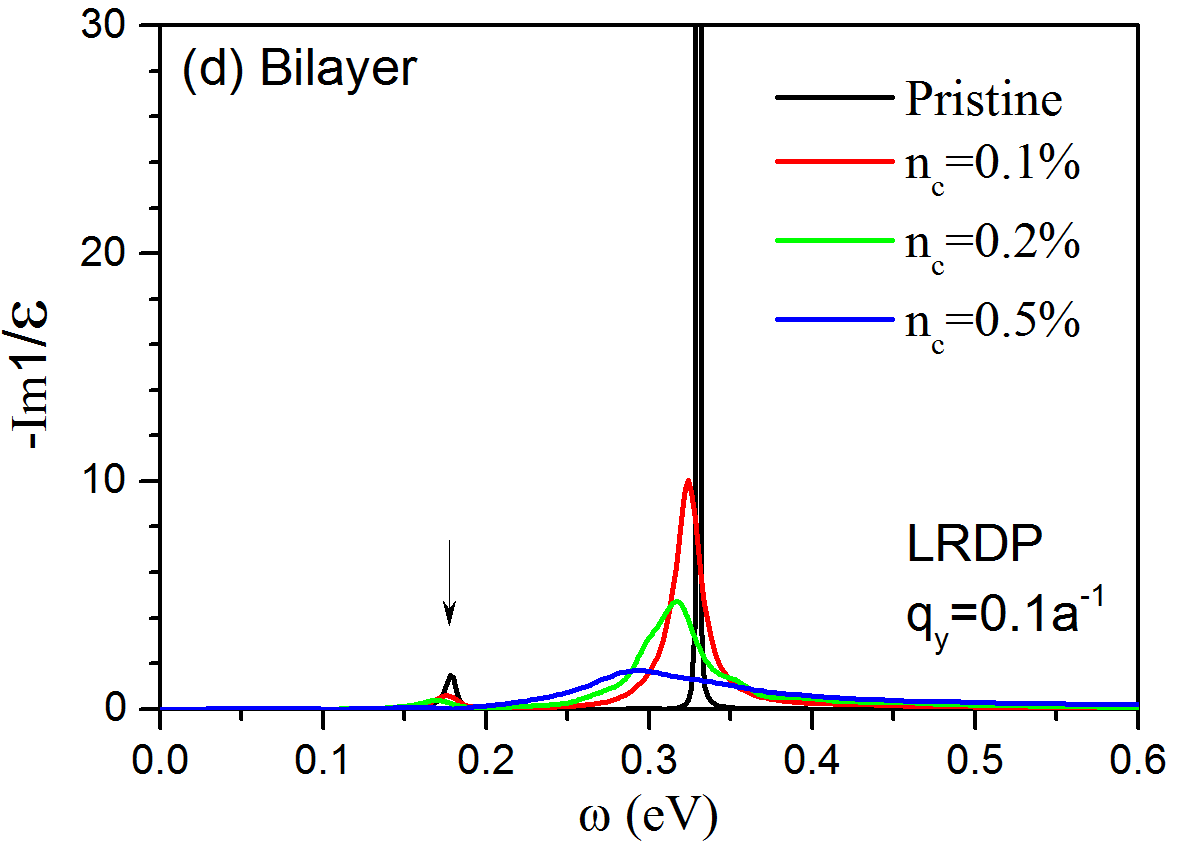}
}
\end{center}
\caption{Energy loss functions of bilayer BP along zigzag (left) and
armchair (right) directions. Panels (a) and (b) compare the loss function of
pristine bilayer BP with that of samples with different concentrations of
point defects. Panels (c) and (d) correspond to LRDP. All the results are
calculated at $T=300$~K. The peak marked by arrows in each panel correspond
to the out-of-phase $\protect\omega_-(q)\sim q$ plasmon (see text).}
\label{Fig:EnergyLossBilayer-q}
\end{figure}

In the following we present results for bilayer BP, considered as two
phosphorene layers coupled by interlayer hopping (see Fig. \ref{Fig:Lattice}%
), and interacting via long range Coulomb potential, as explained in Sec. %
\ref{Sec:Method}. Our results for the loss function are shown in Fig. \ref%
{Fig:EnergyLossbilayer}. The first thing to notice is the existence of two
collective modes in the spectrum, as it is usual in bilayer systems.\cite%
{DasSarma98,G11,Roldan13,Rodin2015} One is the classical $%
\omega_+(q\rightarrow 0)\sim\sqrt{q}$ plasmon mode, which has its
counterpart in single-layer samples (see Fig. \ref{Fig:EnergyLossSingle}).
This mode correspond to a collective excitation of the electron liquid in
which the carriers of both layers oscillate in-phase. In addition to this
plasmon, in Fig. \ref{Fig:EnergyLossbilayer} we observe the existence of an
additional mode in the excitation spectrum dispersing \textit{below} the $%
\omega_+$ plasmon. This mode, with a low energy linear dispersion relation, $%
\omega_-(q\rightarrow 0)\sim q$, is characteristic of bilayer 2DEG systems
and corresponds to a collective oscillation in which the carrier density in
the two layers oscillates out-of-phase. The existence of this mode is also
clear in Fig. \ref{Fig:EnergyLossBilayer-q}, where it can be seen that,
besides the peak in the loss function corresponding to the in-phase $%
\omega_+ $ plasmon, there is a second resonance at lower energies (with less
spectral weight) which corresponds to the $\omega_-\sim q$ plasmon
(indicated by black arrows). The $\omega_-$ mode is much weaker as compared
to the $\omega_+$ mode because of the short distance between the two layers,
which complicates the out-of-phase density oscillation. In Ref. %
\onlinecite{Rodin2015}, the two modes discussed above have been also
obtained by using a general low energy model of anisotropic double-layer
systems. We notice here that, since they study bilayer systems in which the
two layers are separated by more than 5 nm,\cite{Rodin2015} their results
show a $\omega_-$ mode that is more coherent and dispersing than the one
obtained in our calculations. As a matter of fact, the dispersion relation
of the out-of-phase mode is proportional to the layer separation, $%
\omega_-(q)\sim d^{1/2}q$, where $d$ is the interlayer distance. Here we
restrict our study to \textit{real} bilayer phosphorene, in which the two
layers are separated by $d=0.524$~nm, and interlayer hopping of electrons
is allowed. As a consequence, the $\omega_-$ mode obtained in our
simulations for bilayer BP disperses with a smaller velocity than the mode
of Ref. \onlinecite{Rodin2015}.

We also observe that, as in the single-layer case, both modes are damped due
to disorder. The characteristics of the damping of the in-phase $%
\omega_+\sim \sqrt{q}$ mode can be understood from the previous analysis for
single-layer BP. Losses are stronger in the out-of-phase $\omega_-$ plasmon,
whose coherence basically vanishes due to disorder. The reasons for this are
twofold. On the one hand, the spectral weight of the $\omega_-$ mode is
considerably weaker than the $\omega_+$ plasmon. On the other hand, the $%
\omega_-$ mode disperses in the $\omega-q$ spectrum very close to the
electron-hole continuum, whose threshold can be reached due to disorder (and
thermal) broadening of the plasmon peak, with the consequent Landau damping
of the mode.

\subsection{Biased Bilayer BP}

\begin{figure}[t]
\begin{center}
\mbox{
\includegraphics[width=4.25cm]{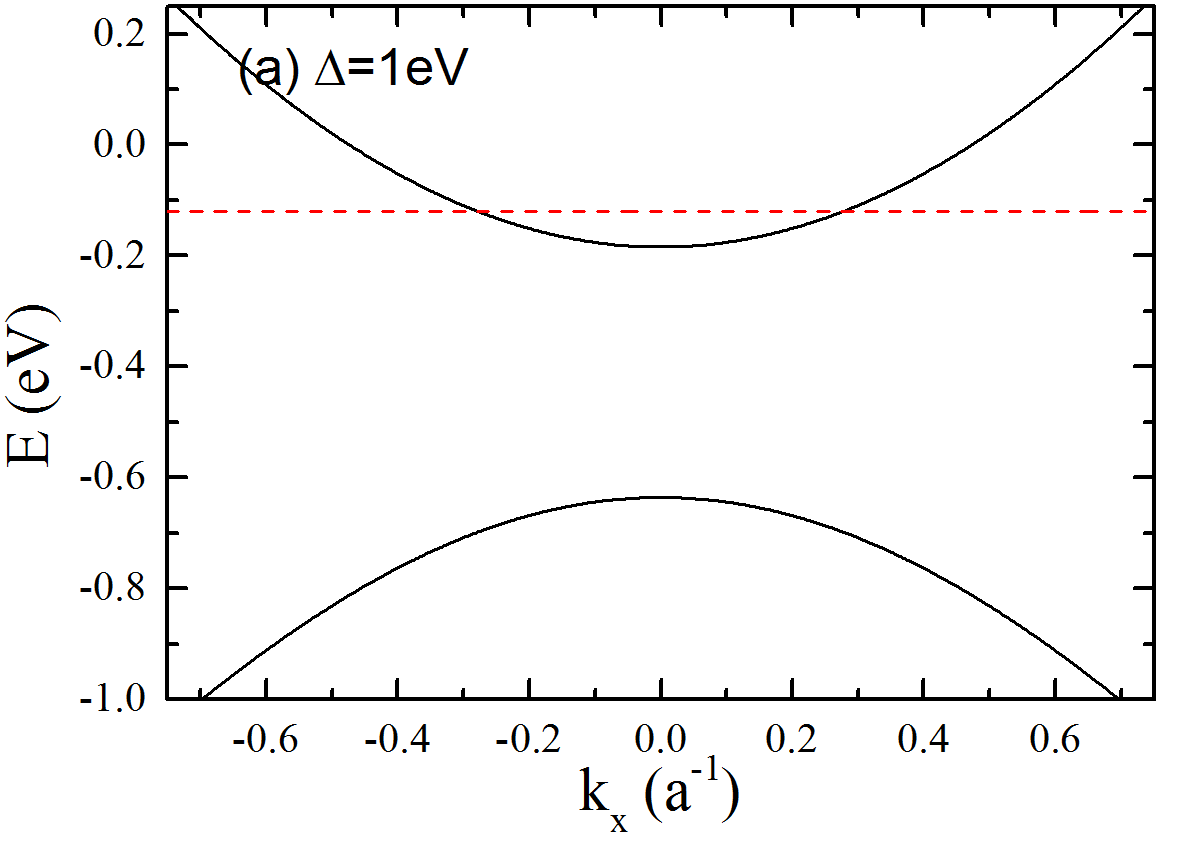}
\includegraphics[width=4.25cm]{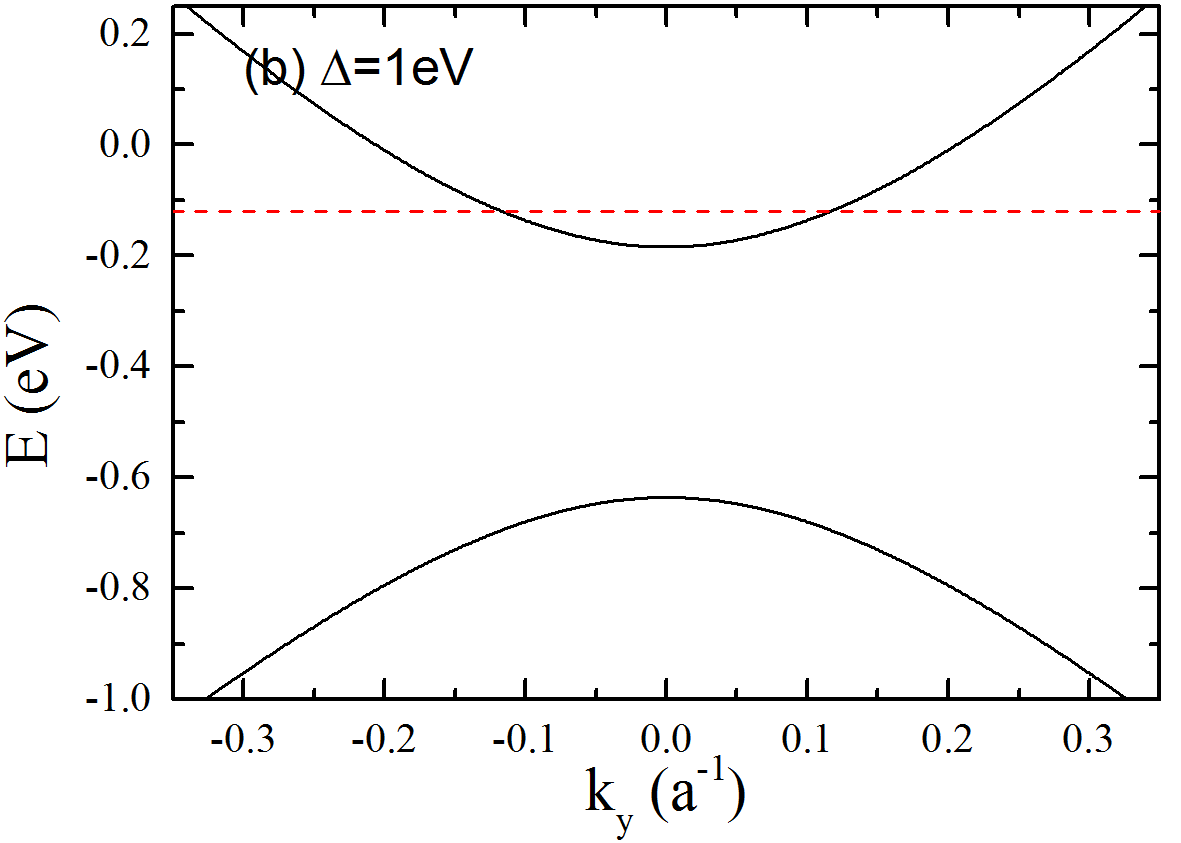}
} 
\mbox{
\includegraphics[width=4.25cm]{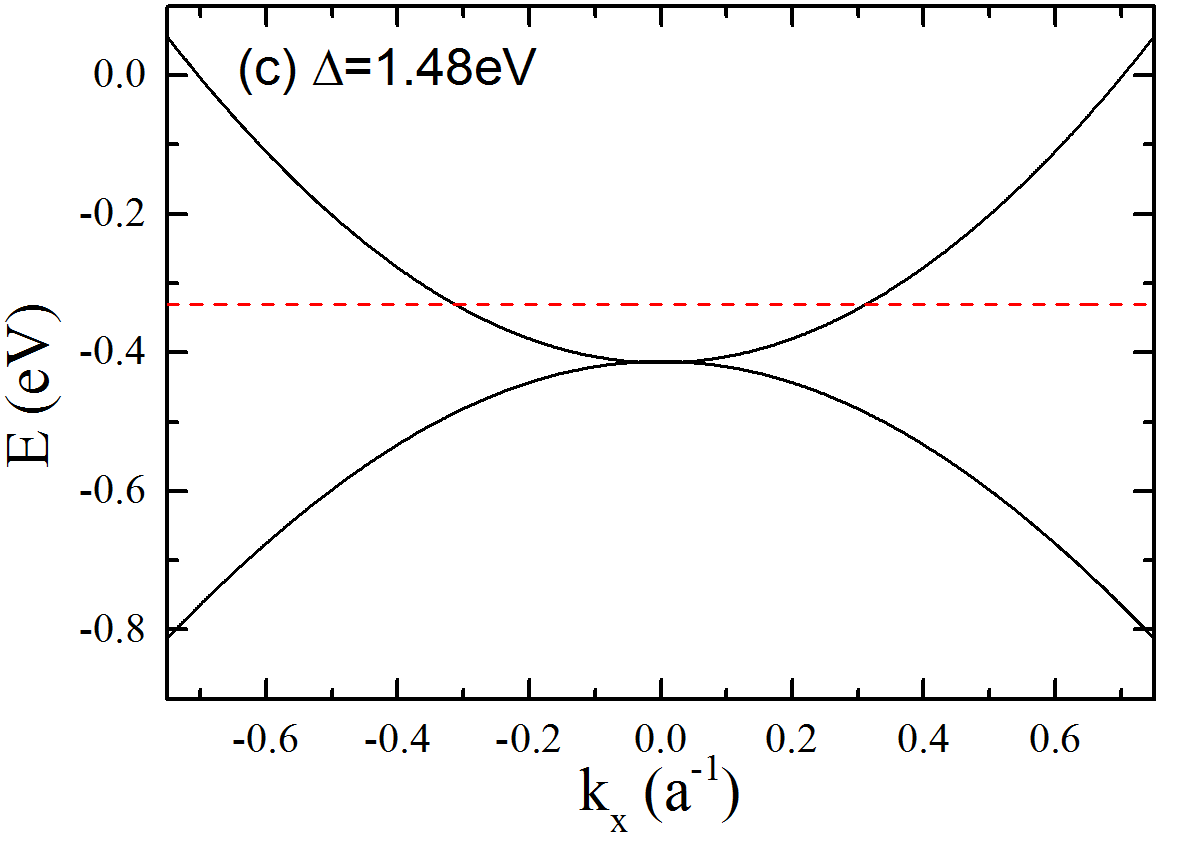}
\includegraphics[width=4.25cm]{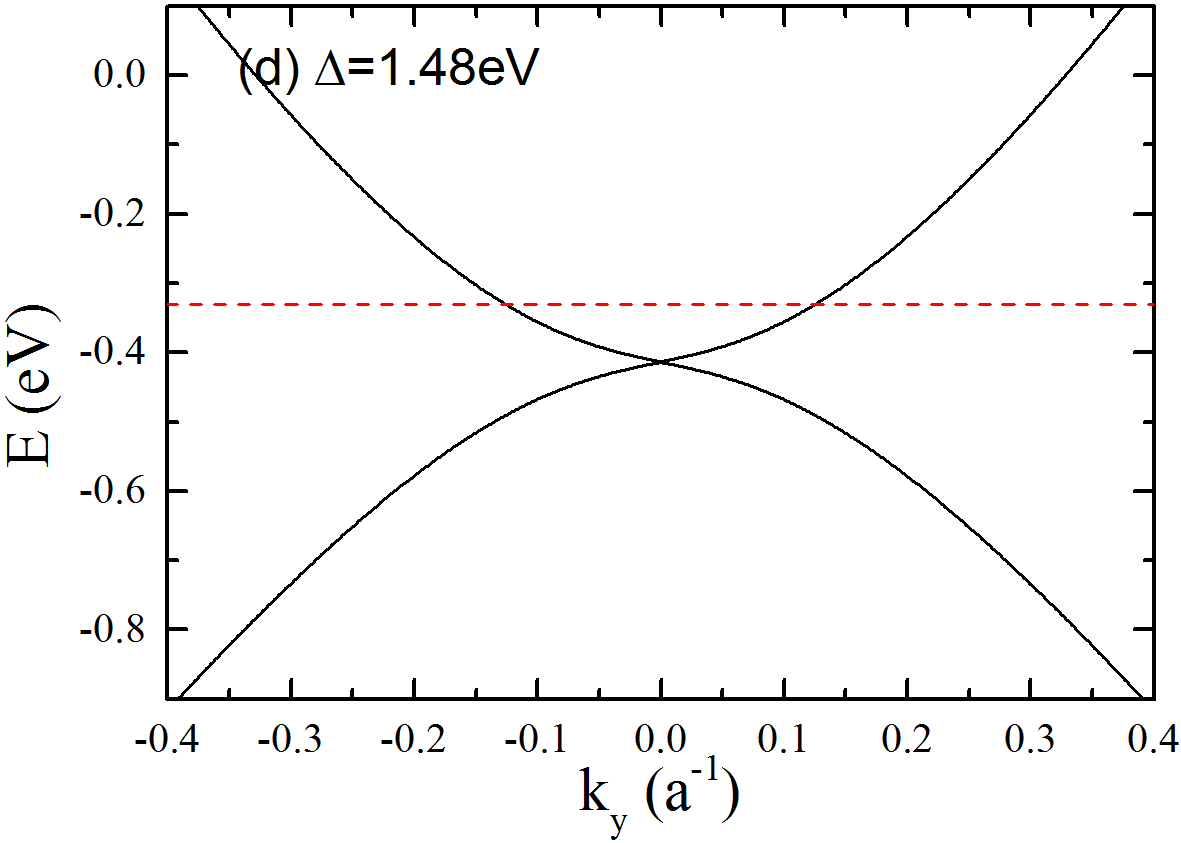}
} 
\mbox{
\includegraphics[width=4.25cm]{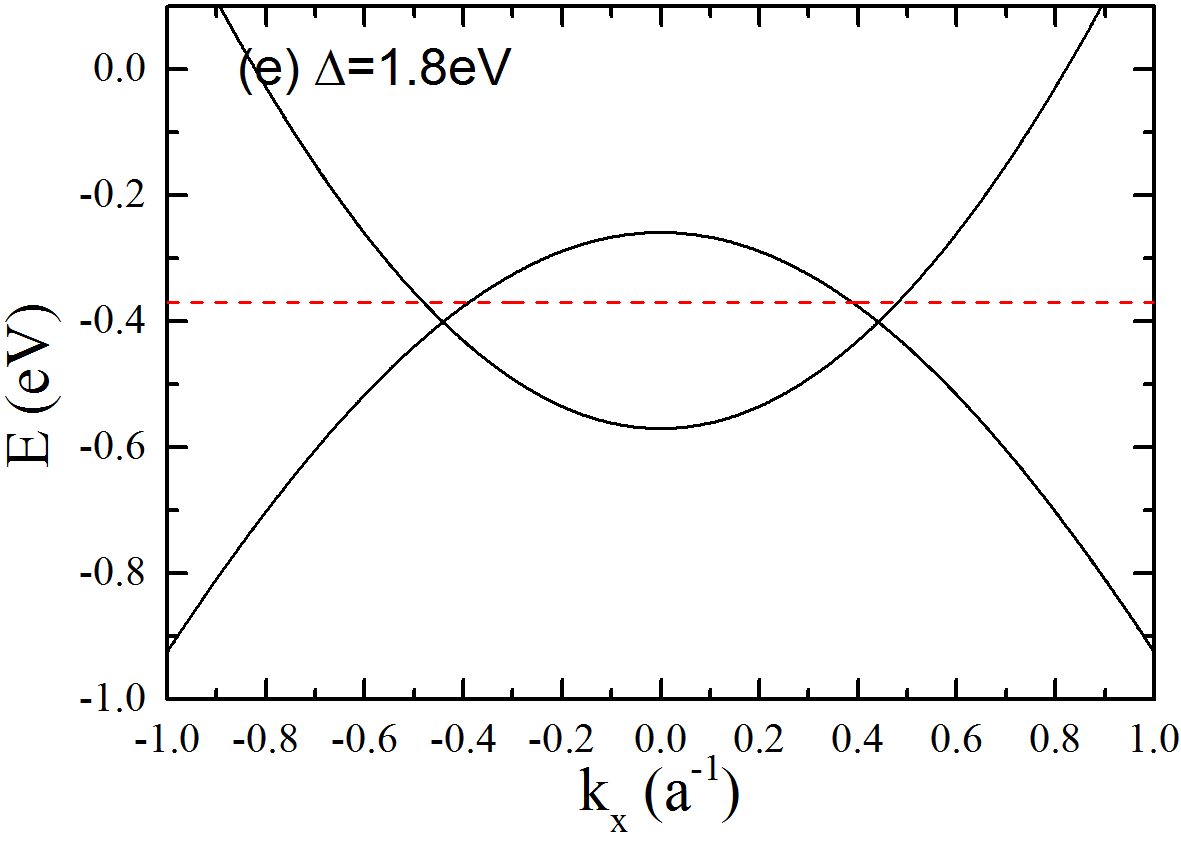}
\includegraphics[width=4.25cm]{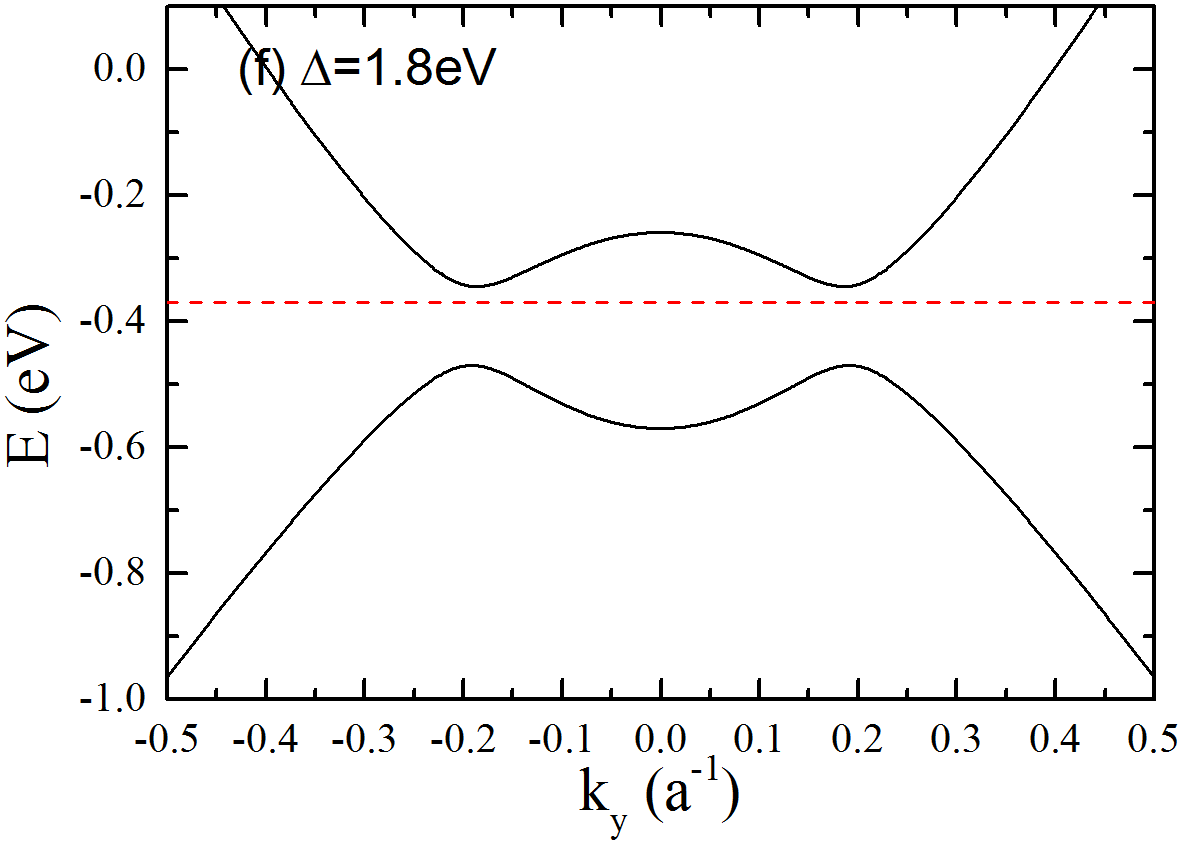}
}
\end{center}
\caption{Low energy band structure of bilayer BP in the presence of a
perpendicular electric field, for different biased potential energy between
top and bottom layers. The red dashed line in each panel indicates the
position of the chemical potential in the calculation of the plasmon modes.}
\label{Fig:BandsBiased}
\end{figure}

In this section we include in our calculations the presence of a
perpendicular electric field applied to the bilayer BP. This technique can
be used to manipulate the electronic band structure of the system, and it
has been proposed recently as an appropriate way to drive a normal to
topological phase in this material.\cite{LZ15,DQ15} Here we are
interested on studying the effect of an electric field on the dispersion
relation of plasmons. For this aim, we introduce a biased on-site potential
difference $\Delta $ between the two layers as described in Ref. %
\onlinecite{Pereira2015}, and consider three representative cases, as shown
in Fig. \ref{Fig:BandsBiased}: $\Delta =1$~eV, for which the band gap is
reduced, $\Delta =1.4$~eV, for which the gap completely closes, and $\Delta
=1.8$~eV, for which the conduction and valence bands are overlapped,
corresponding to the topological phase discussed in Refs. %
\onlinecite{LZ15,DQ15}. It is interesting to notice that, for this last
case, the band structure of bilayer BP presents Dirac like cones for the
dispersion in the $k_{x}$ direction [Fig. \ref{Fig:BandsBiased}(e)], and the
spectrum is gapped in the $k_{y}$ direction [Fig. \ref{Fig:BandsBiased}(f)].
For $\Delta =1$~eV and $1.4$~eV, the chemical potential is chosen to be
about $0.1$~eV above the edge of the conduction band, to better compare with
the results of unbiased bilayer BP of Fig. \ref{Fig:EnergyLossbilayer}. 

\begin{figure}[tb]
\begin{center}
\mbox{
\includegraphics[width=4.25cm]{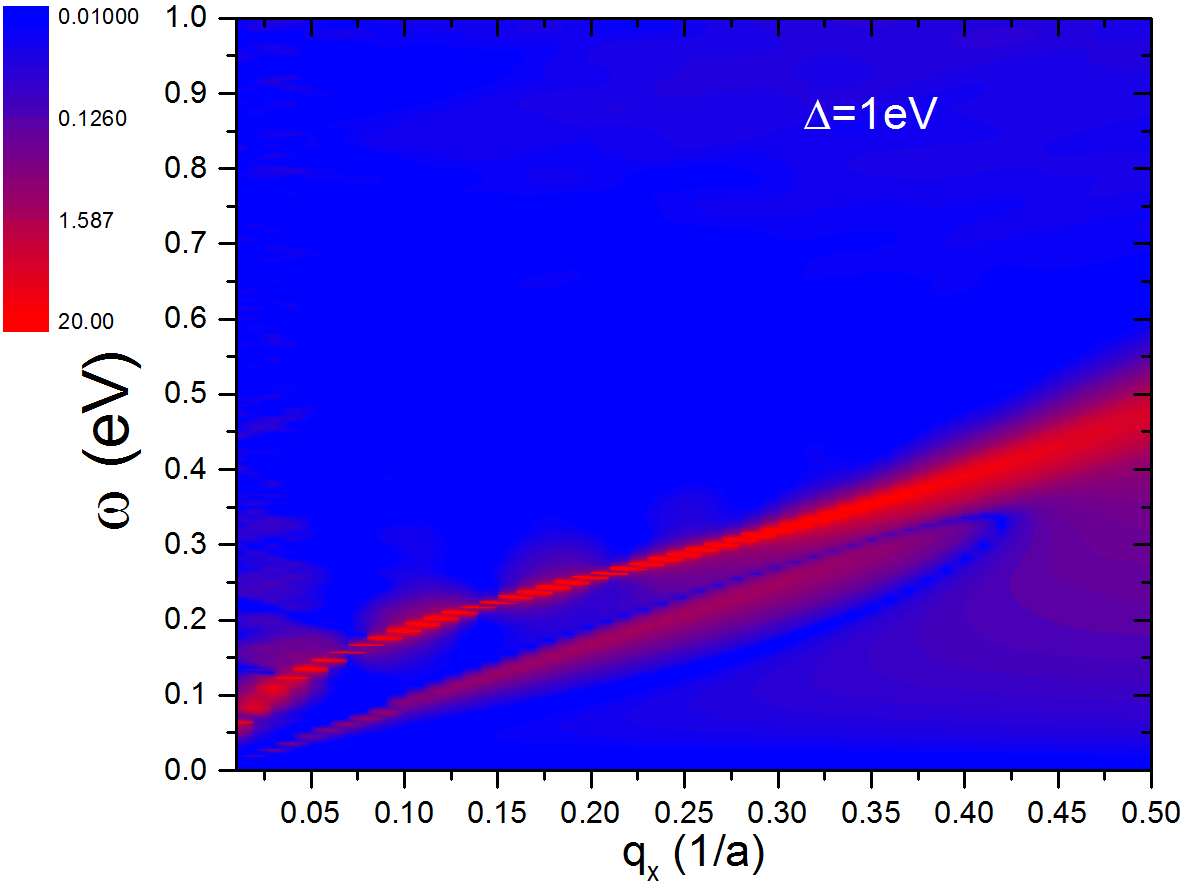}
\includegraphics[width=4.25cm]{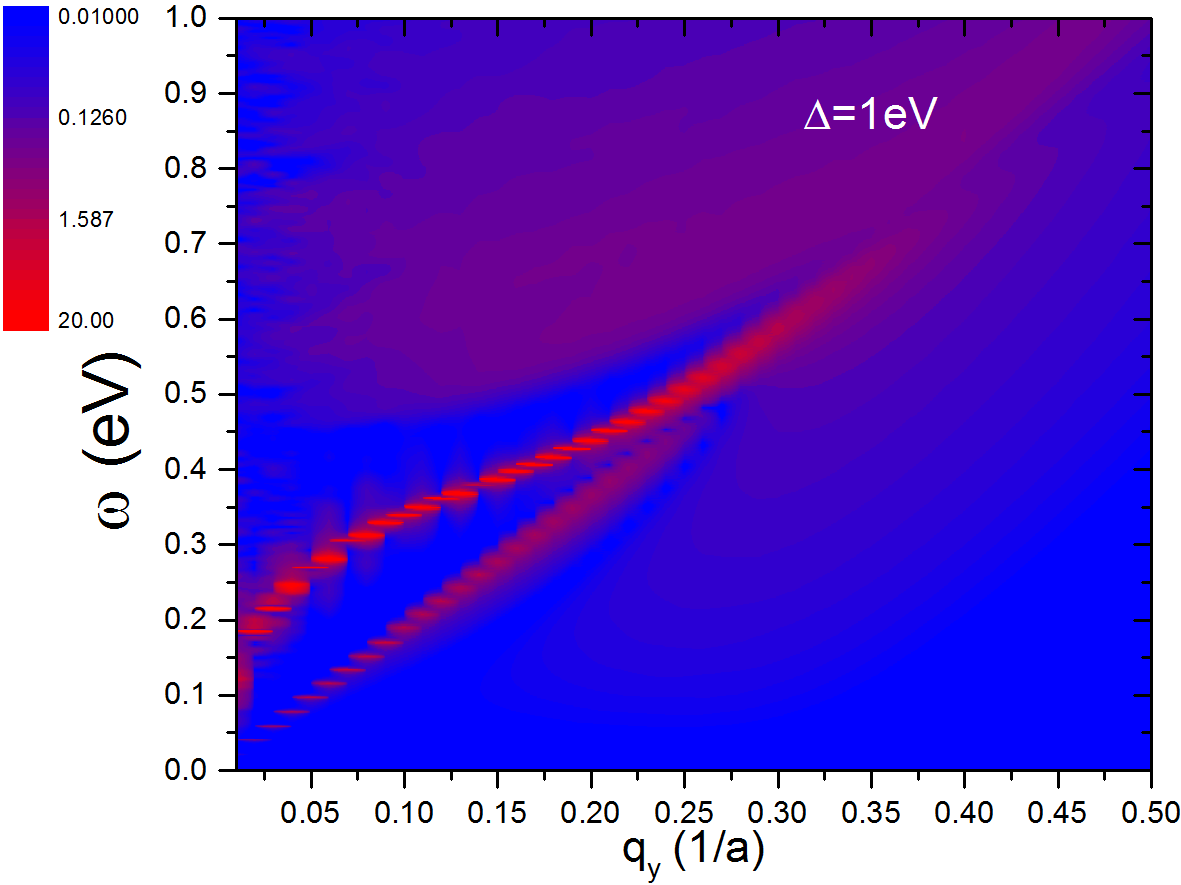}
} 
\mbox{
\includegraphics[width=4.25cm]{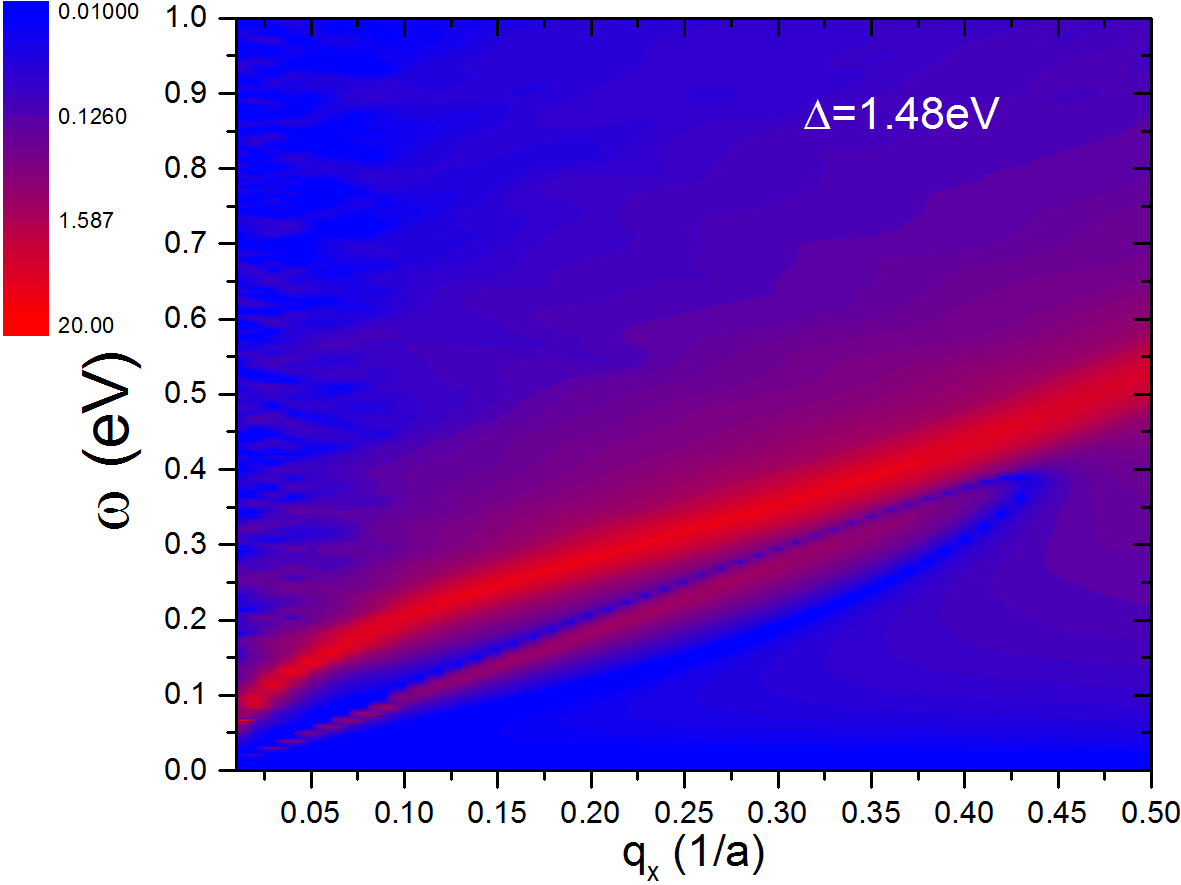}
\includegraphics[width=4.25cm]{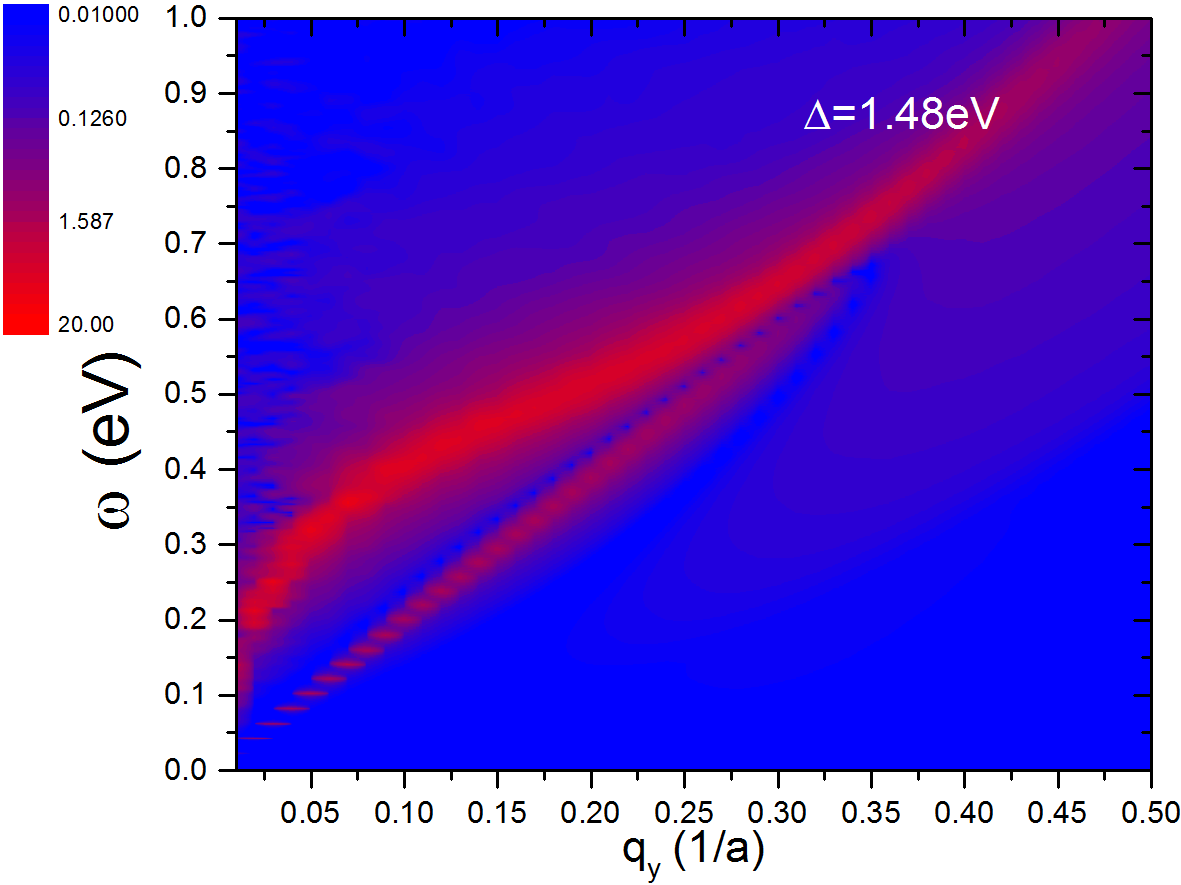}
} 
\mbox{
\includegraphics[width=4.25cm]{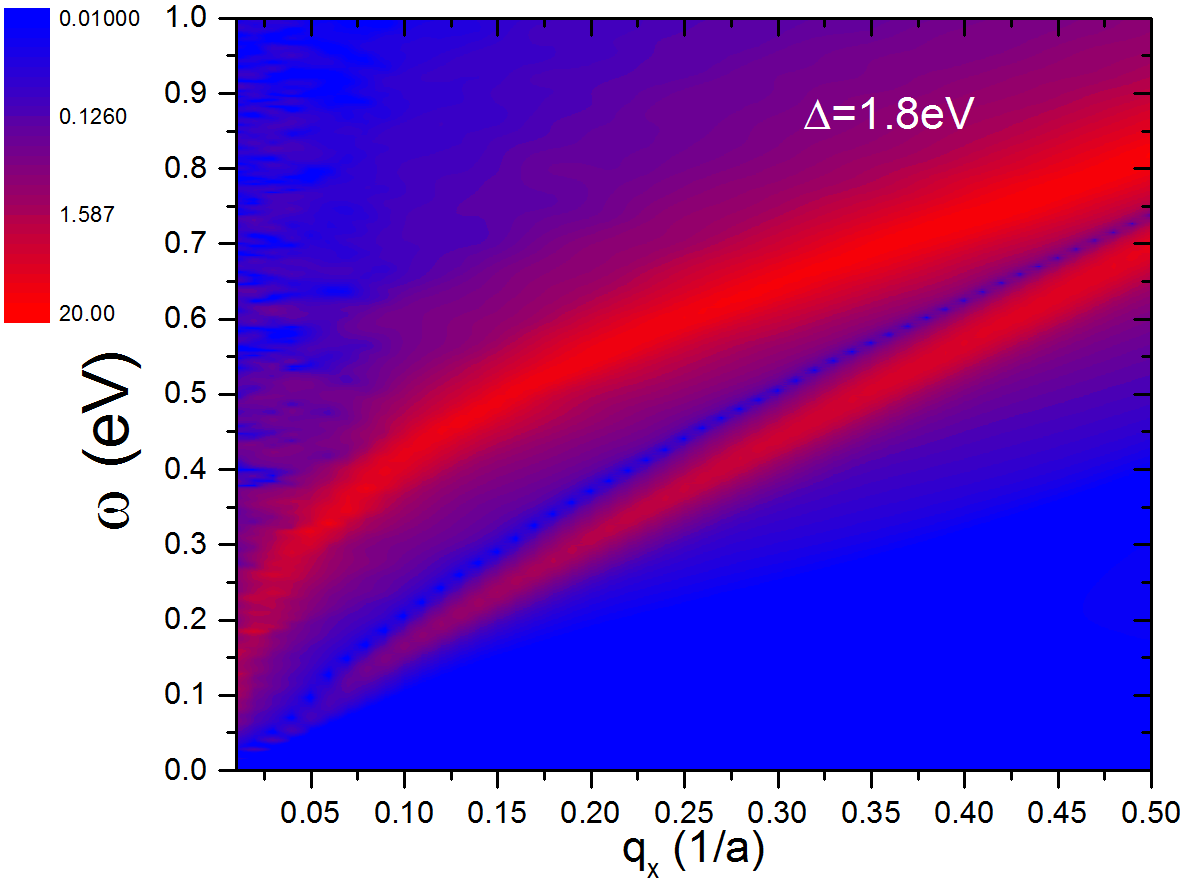}
\includegraphics[width=4.25cm]{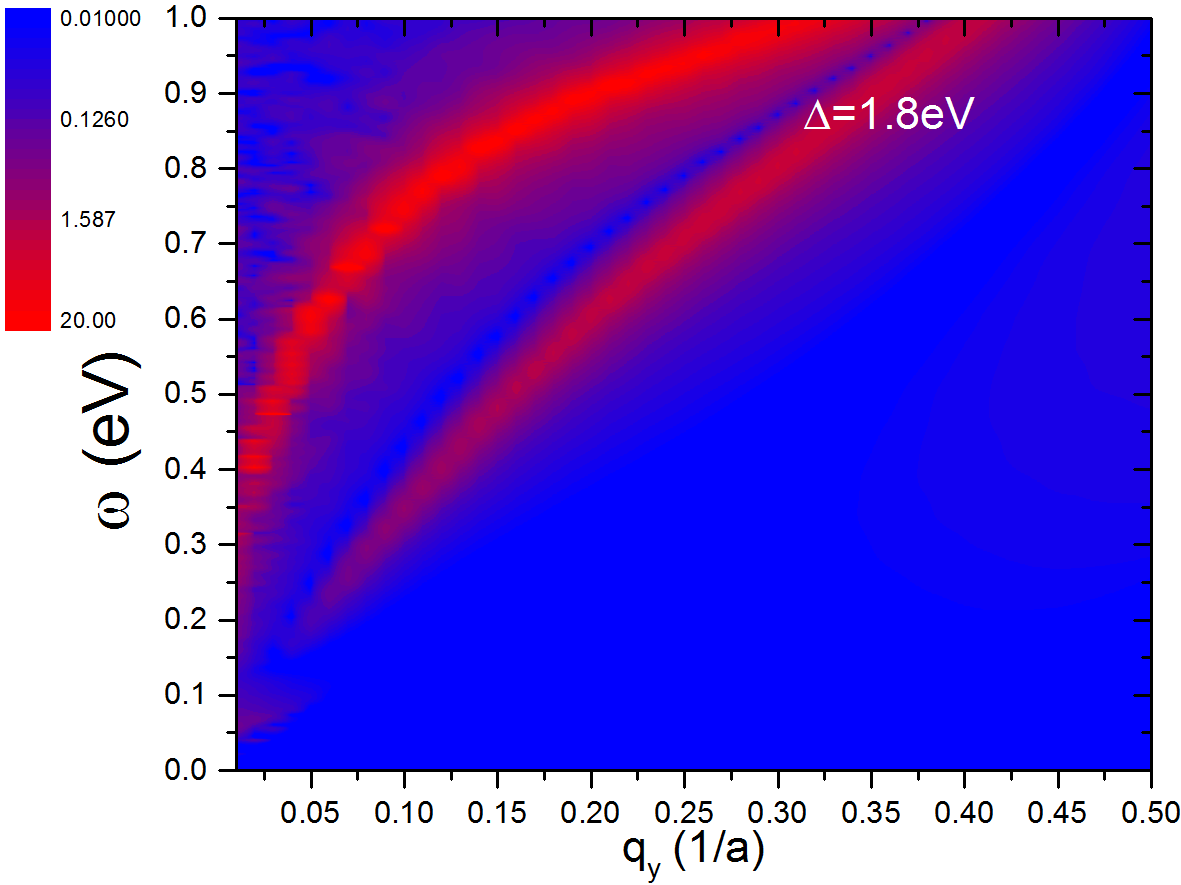}
} 
\mbox{
\includegraphics[width=4.25cm]{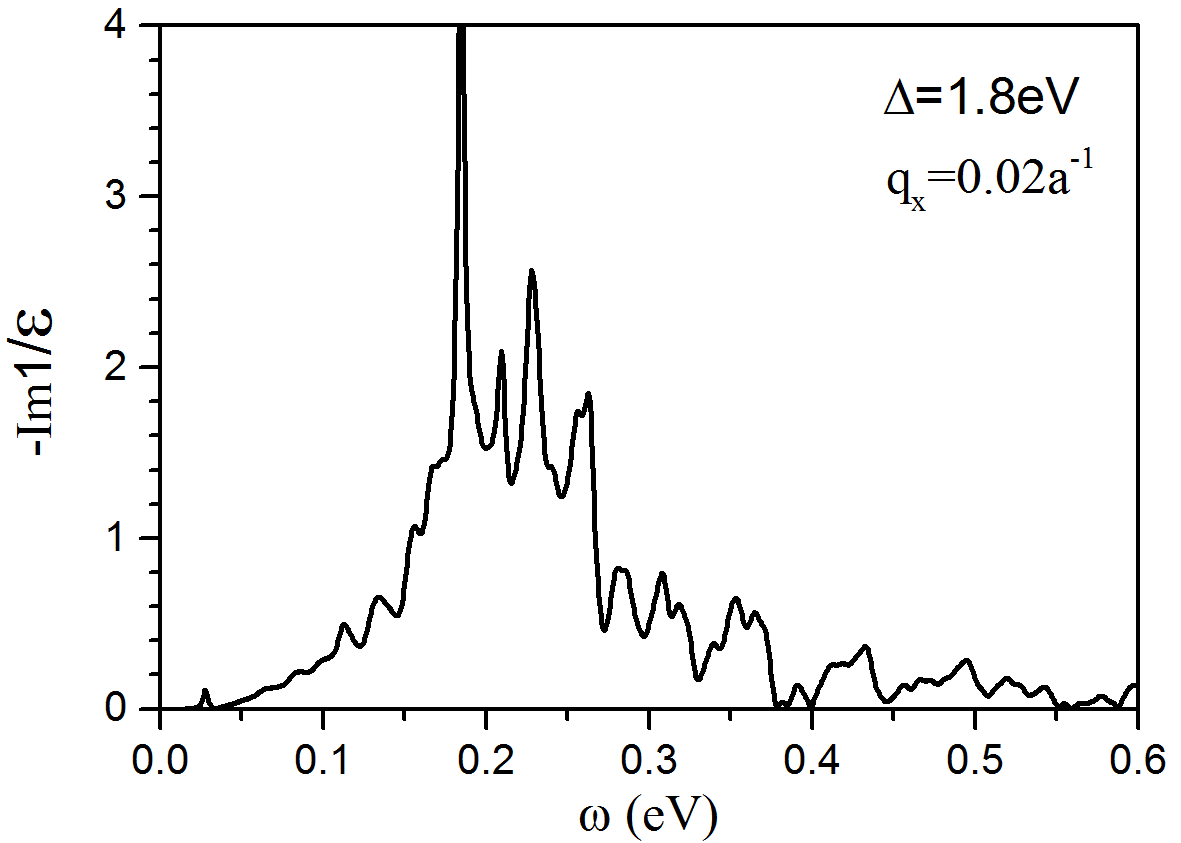}
\includegraphics[width=4.25cm]{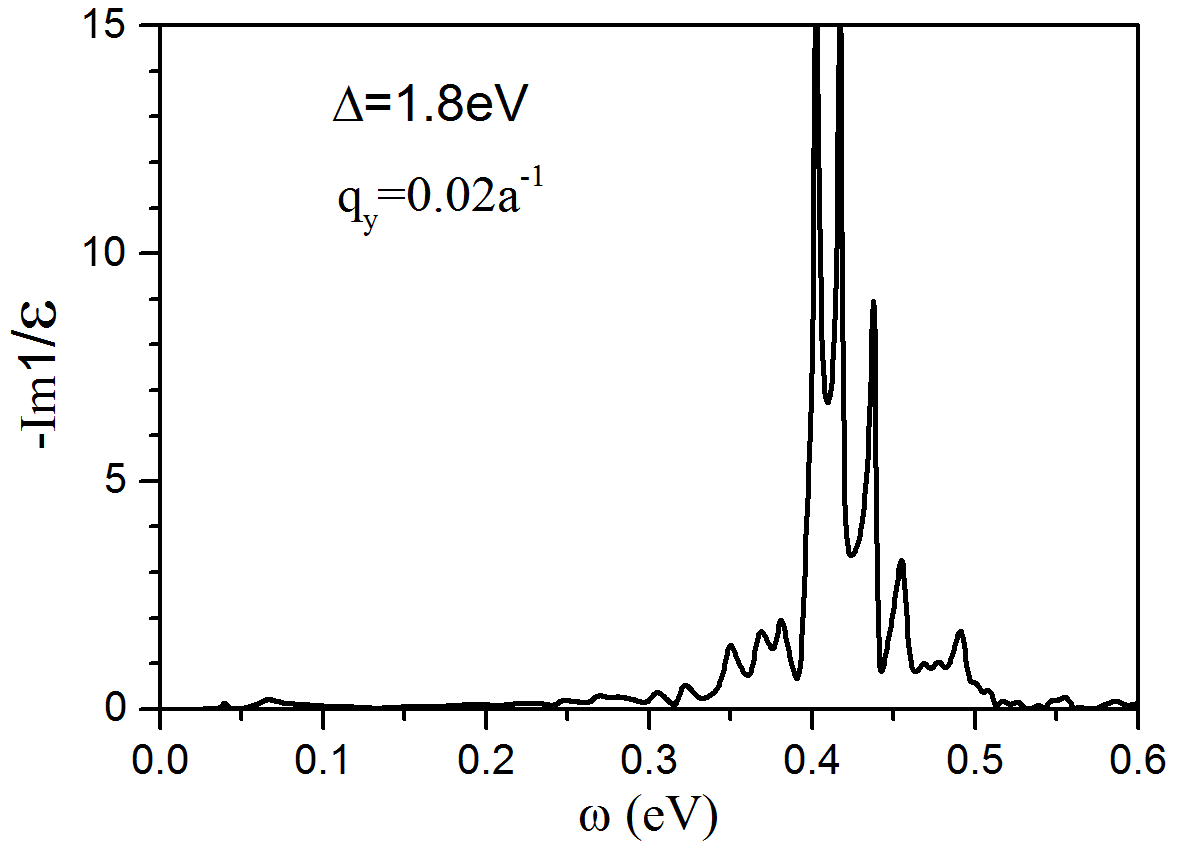}
}
\end{center}
\caption{Energy loss functions of biased bilayer BP along zigzag (left) and
armchair (right) directions. The strength of the biased potential is given
in each panel.}
\label{Fig:EnergyLossBilayerBiased}
\end{figure}

Our results show that the velocity of the plasmon mode increases with the
biased voltage, as it can be seen from the evolution of the excitation
spectrum with the applied $\Delta$ in Fig. \ref{Fig:EnergyLossBilayerBiased}%
. This suggest the possibility to tune the plasmonic properties of this
material with a perpendicular electric field. Interestingly, when the
applied electric field exceeds the critical field and enters in the
topological phase, the plasmons are more coherent, as it can be inferred by
looking at the strength of the modes in the density plots of Fig. \ref%
{Fig:EnergyLossBilayerBiased} (notice the different scales in the density
bar of each panel). Furthermore, the plasmon dispersion is ungapped along $%
q_x$ direction, whereas the presence of a gap in the electronic band
structure in the $\Gamma-$Y direction leads to a gapped collective mode
along $q_y$, as it can be seen by looking Fig. \ref%
{Fig:EnergyLossBilayerBiased}(f) in the low energy and small wave-vector
region. We notice again that, due to the finite size nature of our
simulations, our results cannot reach the $q\rightarrow 0$ limit, for which
we would need to consider an infinite sample. However, the existence of a
gap in the plasmon spectrum $q_y$ direction is already clear in the present
calculation for $\Delta=1.8$~eV. Notice that the peculiar band structure in
the topological phase, with Dirac like band crossings in one direction and
gap in the other direction, leads to a rich excitation spectrum with a
numerous peaks corresponding to the plasmon collective excitations and to
the enhanced optical transitions due to appearance of Van Hove singularities
in the spectrum.

\section{Conclusions}\label{Sec:Conclusions}

In conclusion, we have studied the effect of disorder in the excitation
spectrum of single-layer and bilayer BP. The band structure has been
calculated with an accurate tight-binding model which is valid within $\sim
0.3$~eV beyond the gap. The dynamical polarization function has been
calculated with the Kubo formula, and from this, the energy loss function
has been obtained within the RPA. We have found that disorder leads to a
redshift of the plasmon resonance. This effect has been discussed from a
perturbative point of view, within the framework of the disordered averaged
response function.\cite{Mermin70}

The different kinds of disordered that have been analyzed show that, for the
same concentration of impurities, LRPD leads to stronger damping rates as
compared to local point defects. Such effect is understood from the
intrinsic long wavelength nature of the plasmon oscillations, induced by
long range Coulomb interaction. Therefore resonant scatterers as point
vacancies, which modify the electronic properties at very short length
scales, are less effective inducing losses of the plasmon modes than LRPD.

The spectrum of bilayer BP presents two plasmon modes. One mode in which the
carrier density in the two planes oscillates in-phase, with a dispersion $%
\omega_+(q)\sim \sqrt{q}$, which is the counterpart of the standard plasmon
in single-layer BP, and one mode in which the carriers in the two layers
oscillate out-of-phase, with a liner dispersion relation at low energies, $%
\omega_-(q)\sim q$. The coherence of the $\omega_-$ mode is more
dramatically affected by disorder.

Finally, we have studied the effect of a perpendicular electric field in the
excitation spectrum of bilayer BP. We have shown that the dispersion of the
collective modes can be tuned by the application of such biased field.
Furthermore, we have shown that, beyond some critical field, the bilayer BP
enters in a topological phase with Dirac like crossing of bands in one
direction and gapped in the other direction. As a consequence, the
excitation spectrum is highly rich in this limit, with highly coherent
plasmon modes which are gapped in the $q_y$ direction of the spectrum. In
summary, our results show a highly anisotropic excitation spectrum of single
layer and bilayer BP, features that could be of high interest for future
optoelectronic applications.

\section{Acknowledgments}

The support by the Stichting Fundamenteel Onderzoek der Materie (FOM) and
the Netherlands National Computing Facilities foundation (NCF) are
acknowledged. S.Y. and M.I.K. thank financial support from the European
Research Council Advanced Grant program (contract 338957). The research has
also received funding from the European Union Seventh Framework Programme
under Grant Agreement No. 604391 Graphene Flagship. R.R. acknowledges support
from the European Research Council Advanced Grant (contract 290846).

\bibliographystyle{apsrev}
\bibliography{Bib_2D}

\end{document}